\documentclass[12pt]{report}
\usepackage[utf8]{inputenc}
\usepackage{blindtext}
\usepackage[letterpaper, total={8.5in, 11in}, margin=2in]{geometry}
\usepackage{ragged2e}
\usepackage{blindtext}
\usepackage{graphicx}
\usepackage{amsmath}
\usepackage{amssymb}
\usepackage{gensymb}
\usepackage{array}
\date{}

\begin{document}
\centering{\huge Maximization of User Association Deploying IRS in 6G Networks\\
\vspace{24pt}
\large Mobasshir Mahbub, Raed M. Shubair}

\newpage

\RaggedRight{\textbf{\Large 1.\hspace{10pt} Introduction}}\\
\vspace{18pt}
\justifying{\noindent Connectivity technologies and processes, as well as the services that use them, have undergone exponential development and tremendous progress throughout the last decades. Examples include indoor localisation and related characteristics [1-39], terahertz transmitting and signal processing strategies [40-65], and antenna architecture and propagating properties [66-100].

The aims of sixth-generation (6G) wireless networks are projected to be disruptive and revolutionary, including applications such as data-driven, rapid, ultra-massive, and pervasive wireless networking [101], [102]. As a result, new transmission technologies are required to enable these emerging services and applications.

Emerging wireless systems will put a load on the existing single-tier macro cell-based networks to meet rising mobile data traffic demands. Currently, academic and industrial experts have presented a viable approach for mobile networks in which an increasing number of low-cost, low-power consuming, and short-extent tiny base stations (BSs) are installed under the traditional macro BSs. This advancement transforms standard single-tier communication networks into multi-tier networks. The -refore, the multi-tier network design appears as one of the potential solutions to the evolving traffic growth problem.
\begin{figure}[htbp]
\centerline{\includegraphics[height=6.5cm, width=10.5cm]{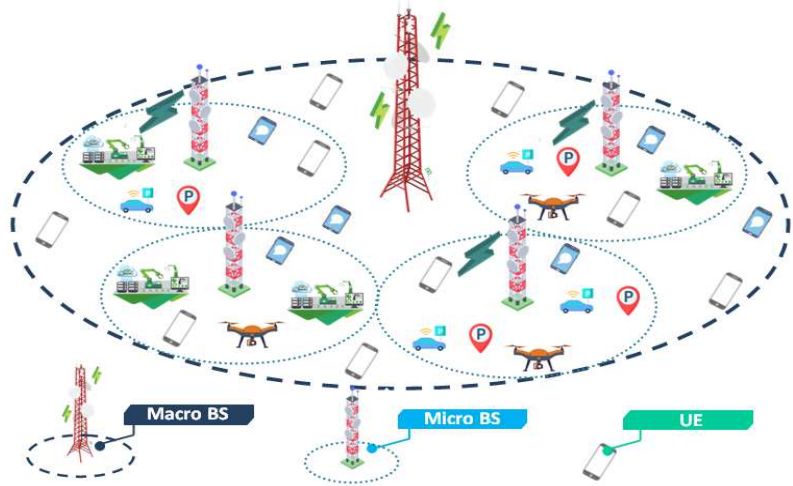}}
\caption{A typical two-tier network.}
\label{fig}
\end{figure}
Various low-power base stations (e.g., micro, pico, and femto base stations) are implemented in a multi-tier network to alleviate macro cell users, establishing a multi-tier network layered with numerous tiny cells [103]. Fig. 1 visualizes a typical multi-tier (two-tier) network.

IRS also termed as RIS, large intelligent surfaces (LIS), software -controlled metasurfaces  [104], [105] has recently emerged as a new potential solution for improving the functioning of wireless systems. IRS may automatically modify the wireless transmission environment for improved directional signaling. IRS is similar to a full-duplex amplify-and-forward (AF) relay that is composed of a large quantity of passive reflecting components that may forward incident signals via passive beamforming [106]. Furthermore, IRS can modify propagation channels dynamically based on real environmental circumstances by fabricating phase shifts of the incoming signals. As a result, the IRS component cost and energy consumption are significantly lower compared to the AF relay. Research derived that the deployment of IRSs can highly enhance the evolving network improvement schemes such as (unmanned aerial vehicle) UAV-assisted communication [107], backscatter communication [108], etc. Moreover, IRS can facilitate application scenarios such as vehicle-to-everything (V2X) services [109], multi-access edge computing (MEC) [110], etc.

The extensive implementation of small base stations pushes network access points nearer to user devices (UD) or equipment (UE), resulting in improved traffic dispatching from the macro cell base station to the small cell base station. However, to fully realize its potential, several crucial challenges must be solved. One of the most significant issues is user association [111], which entails deciding which base station the device should be linked to.

Although IRS can provide numerous benefits to the mmWave transmission systems, it is important to note that IRS has a substantial influence on user associations. The received power strength-based user association technique is commonly employed in mmWave communications, which may result in wasteful resource consumption (e.g. transmit power of the BS). There are several research initiatives underway to increase system performance by taking into account issues such as fairness of coverage and load balancing in mmWave transmission networks. The study's goal is to increase network utility by maximizing user association.

Therefore, the target of this research is to maximize the user association for mmWave carriers in a two-tier 6G network. The work deployed IRS in a micro cell to maximize or enhance the probability of user association and compared it with a conventional non-IRS micro cell.
}
\vspace{18pt}

\RaggedRight{\textbf{\Large 2.\hspace{10pt} Related Literature}}\\
\vspace{18pt}
\justifying{\noindent The section briefed several prior research works relative to IRS-assisted multi-tier networks.

Hashida et al. [112] proposed and analyzed an IRS–assisted user association scheme considering the mobility of devices in the context of multi-beam (IRS-aided) transmissions for 6G networks. Zhao et al. [113] analyzed the IRS-assisted user association in mmWave-based multi-base station 5G systems. The work maximized the sum rate through the joint optimization of user association, passive beamforming, and power allocation. Taghavi et al. [114] proposed a novel user association scheme for load balancing for mmWave-based 5G and beyond transmission networks where IRS is implemented to escalate the coverage. Mei et al. [115] considered an IRS-assisted multi-tier (base stations) network and investigated user association in the case of the downlink. Belaoura et al. [116] analyzed the user association optimization problem for the relay-UAV-based massive multi-input multi-output (mMIMO) mmWave 6G communication systems. However, IRS is not considered in this network. Munir et al. [117] proposed and analyzed dual-slope path loss model-based user association considering a 5G two-tier network.

During the literature review, the authors found a limited number of works related to the user association analysis for 6G networks. The number of literature relative to IRS/RIS-assisted user association is still much lower as well. The notable thing is that the previous articles have considered a single carrier frequency for measurement. This work considered multiple carrier frequencies from a vast range of frequencies (i.e., 28-90 GHz) for measurement and analysis.
}
\vspace{18pt}

\newpage
\RaggedRight{\textbf{\Large 3.\hspace{10pt} Measurement Model}}\\
\vspace{18pt}
\justifying{\noindent Contemplating a two-tier network consisting of a set of micro base stations functioning under a macro base station. The macro and micro base station is serving the corresponding user devices. In the context of an IRS-enhanced micro cell, the micro base station will serve the user through an IRS. $P_t$ is the transmit (or transmission) power of the micro base station.

\vspace{12pt}
\RaggedRight{\textit{\large A.\hspace{10pt} Conventional Model}}\\
\vspace{12pt}

\justifying The downlink received power of a conventional micro base station is measured by the following equation (Eq. 1) [118]-[120],

\begin{equation}
P^{Micro(Conv.)}_{r} = \frac{P_t\lambda^2h}{16\pi^2d^\alpha}
\end{equation}
where $h$ is the Rayleigh fading coefficient that follows unit mean-exponential distribution, i.e., $h\sim exp(1)$. The wavelength is denoted by $\lambda= c⁄f_c$. $c$ is the speed of light in $ms^{-1}$. $f_c$ is the frequency in Hz.

$d= \sqrt{(x^\mathsf{n}-x^\mathsf{u})^2+(y^\mathsf{n}-y^\mathsf{u})^2+(z^\mathsf{n}-z^\mathsf{u})^2}$ denotes the distance between the user located at $(x^\mathsf{u},y^\mathsf{u},z^\mathsf{u})$ and the micro base station placed at $(x^\mathsf{n},y^\mathsf{n},z^\mathsf{n})$ coordinates. $\alpha$ is the attenuation exponent representing the degradation of the transmitted signal.

\vspace{12pt}
\RaggedRight{\textit{\large B.\hspace{10pt} IRS-Assisted Model}}\\
\vspace{12pt}

\justifying The received power in the downlink of an IRS-assisted micro base station can be formulated by (Eq. 2) [121],

\begin{equation}
P^{Micro(IRS)}_{r} = P_t\frac{G_{\mathsf{Sc}} G_t G_r M^2 N^2 d_x d_y \lambda^2 cos\theta_t cos\theta_r A^2}{d^2_1 d^2_2 64\pi^3}
\end{equation}
where $d_x$ and $d_y= \lambda/2$. The IRS elements’ scattering gain is $G_{Sc}=\frac{4\pi d_x d_y}{\lambda^2}$ . The gains of transmitter and receiver are indicated by $G_t$ and $G_r$.
\begin{equation*}
d_1 = \sqrt{(x^{\mathsf{n}}-x^{i})^2+(y^{\mathsf{n}}-y^{i})^2+(z^{\mathsf{n}}-z^{i})^2}
\end{equation*} is the distance between the micro base station at $(x^\mathsf{n},y^\mathsf{n},z^\mathsf{n})$ coordinates and IRS at $(x^i,y^i,z^i)$ coordinates.
\begin{equation*}
d_2 = \sqrt{(x^{i}-x^{\mathsf{u}})^2+(y^{i}-y^{\mathsf{u}})^2+(z^{i}-z^{\mathsf{u}})^2}
\end{equation*} denotes the separation between the IRS $i$ and the user at

$(x^\mathsf{u},y^\mathsf{u},z^\mathsf{u})$ coordinates. The numbers of the IRS’s transmitter and receiver elements are indicated by $M$ and $N$, respectively.  The length and width of the scattering elements are $d_x$ and $d_y$, respectively. $\theta_t$ denotes the angle between the micro base station and IRS and $\theta_r$ is the angle between IRS-to-user equipment. $A$ is the reflection coefficient of the IRS.

\newpage
\vspace{12pt}
\RaggedRight{\textit{\large C.\hspace{10pt} User Association}}\\
\vspace{12pt}

\justifying The work includes the equations to measure and analyze the probability of user association (UA) for both conventional and IRS –assisted micro base stations.

Conventional UA: For a conventional (non-IRS) micro base station the probability of user association is measured by the equation below (Eq. 3) [122],

\begin{equation}
\mathcal{A}^{Micro(Conv.)} = \left (1+\frac{\lambda_{Macro}}{\lambda_{Micro}}\left(\frac{P^{Macro}_r}{P^{Micro(Conv)}_r}\right)^\frac{2}{\alpha_{macro}}\right)^{-1}
\end{equation}
where $\lambda_{Micro}$  and $\lambda_{Macro}$ are the densities (per $m^2$) of the micro base station and macro base station, respectively. $P_r^{Micro(Conv.)}$ and $P_r^{Macro}$ denote the received power from the serving conventional micro base station and the macro base station (by the user), respectively. $\alpha_{macro}$ indicates the exponent representing the attenuation for the macro base station.

IRS-Assisted UA: The probability of user association for an IRS-empowered micro base station is determined by (Eq. 4) [122],

\begin{equation}
\mathcal{A}^{Micro(IRS)} = \left (1+\frac{\lambda_{Macro}}{\lambda_{Micro}}\left(\frac{P^{Macro}_r}{P^{Micro(IRS)}_r}\right)^\frac{2}{\alpha_{macro}}\right)^{-1}
\end{equation}
where $P^{Micro(IRS)}_r$ is the received power from the serving IRS-enhanced micro cell.
}
\vspace{18pt}

\RaggedRight{\textbf{\Large 4.\hspace{10pt} Numerical Results and Discussions}}\\
\vspace{18pt}
\justifying{\noindent The section incorporates the numerical results and corresponding discussions on the simulated results. The work considers that the micro base station is placed at the cell center namely (100, 100) coordinates in the case of a conventional micro cell, and in the case of an IRS-assisted micro cell the IRS is placed at the cell center. Furthermore, the research contemplates that, in the case of an IRS-assisted micro cell the micro base station serves all users through the IRS. Table I represent the simulation parameters and values for the measurements.

\begin{table}[htbp]
\caption{Parameters and Values}
\begin{center}
\begin{tabular}{| m{3.5cm} | m{3.5cm}|}
\hline
\textbf{\textit{Parameters}}& \textbf{\textit{Values}}\\
\hline
Macro cell area & 1000x1000m\\
\hline
Micro cell area & 200x200m\\
\hline
Transmit power of the Macro BS & 50 W \\
\hline
Transmit power of the Micro BS, $P_t$ & 4-6 W (conv.), 1-2 W (IRS-assisted) \\
\hline
BS's position, $(x^\mathsf{n},y^\mathsf{n},z^\mathsf{n})$ & Conv.: (100, 100, 5) and IRS-assisted: (0, 0, 5) coordinates\\
\hline
Users position, $(x^\mathsf{u},y^\mathsf{u},z^\mathsf{u})$ & Randomly distributed\\
\hline
IRS's position, $(x^i,y^i,z^i)$ & (100, 100, 5) coordinates\\
\hline
Transmitter and receiver gain, $G_t$ and $G_r$ & 20 dB [123]\\
\hline
\end{tabular}
\label{tab1}
\end{center}
\end{table}

\begin{table*}[htbp]
\begin{center}
\begin{tabular}{| m{3.5cm} | m{3.5cm}|}
\hline
\textbf{\textit{Parameters}}& \textbf{\textit{Values}}\\
\hline
Transmitter and receiver elements, $M$ and $N$, & 128, 256\\
\hline
Transmit angles, $\theta_t$ & 45$\degree$, 60$\degree$ [121], [124]\\
\hline
Receive angles, $\theta_r$ & 45$\degree$, 60$\degree$ [121], [124]\\
\hline
mmWave carriers, $f_c$ & 28, 50, 70, 90 GHz\\
\hline
Reflection coefficient of the IRS, $A$ & 0.9 [21]\\
\hline
Micro cell tier BS density, $\lambda_{micro}$ & $500⁄(\pi(100)^2 )$ per $m^2$; where 100 m is the cell radius\\
\hline
Macro cell tier BS density, $\lambda_{macro}$ & $\lambda_{micro}⁄5$ per $m^2$\\
\hline
Path loss exponent (Micro cell), $\alpha$ & 2.5\\
\hline
Path loss exponent (Macro cell), $\alpha_{macro}$ & 4.5\\
\hline
\end{tabular}
\label{tab1}
\end{center}
\end{table*}

Fig. 2 (a) and (b) illustrate the probability of user association corresponding to the selected mmWave carriers in the context of a conventional non-IRS micro cell for 4 W and 6 W of transmission (or transmit) powers, respectively.

\begin{figure}[htbp]
\centerline{\includegraphics[height=7.5cm, width=10.5cm]{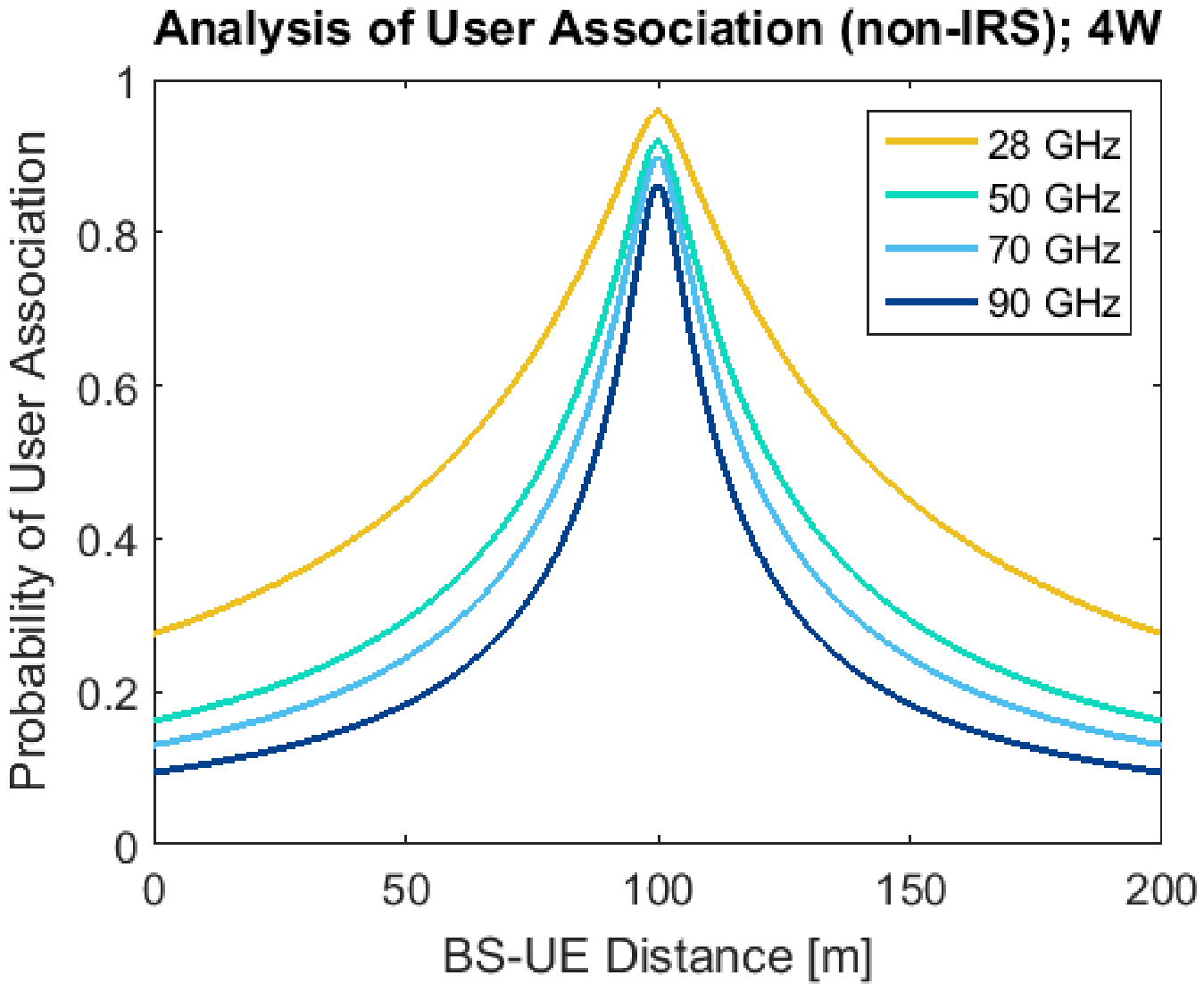}}
\vspace{3pt}
\centerline{\footnotesize{(a)}}
\label{fig}
\end{figure}

\begin{figure}[htbp]
\centerline{\includegraphics[height=7.5cm, width=10.5cm]{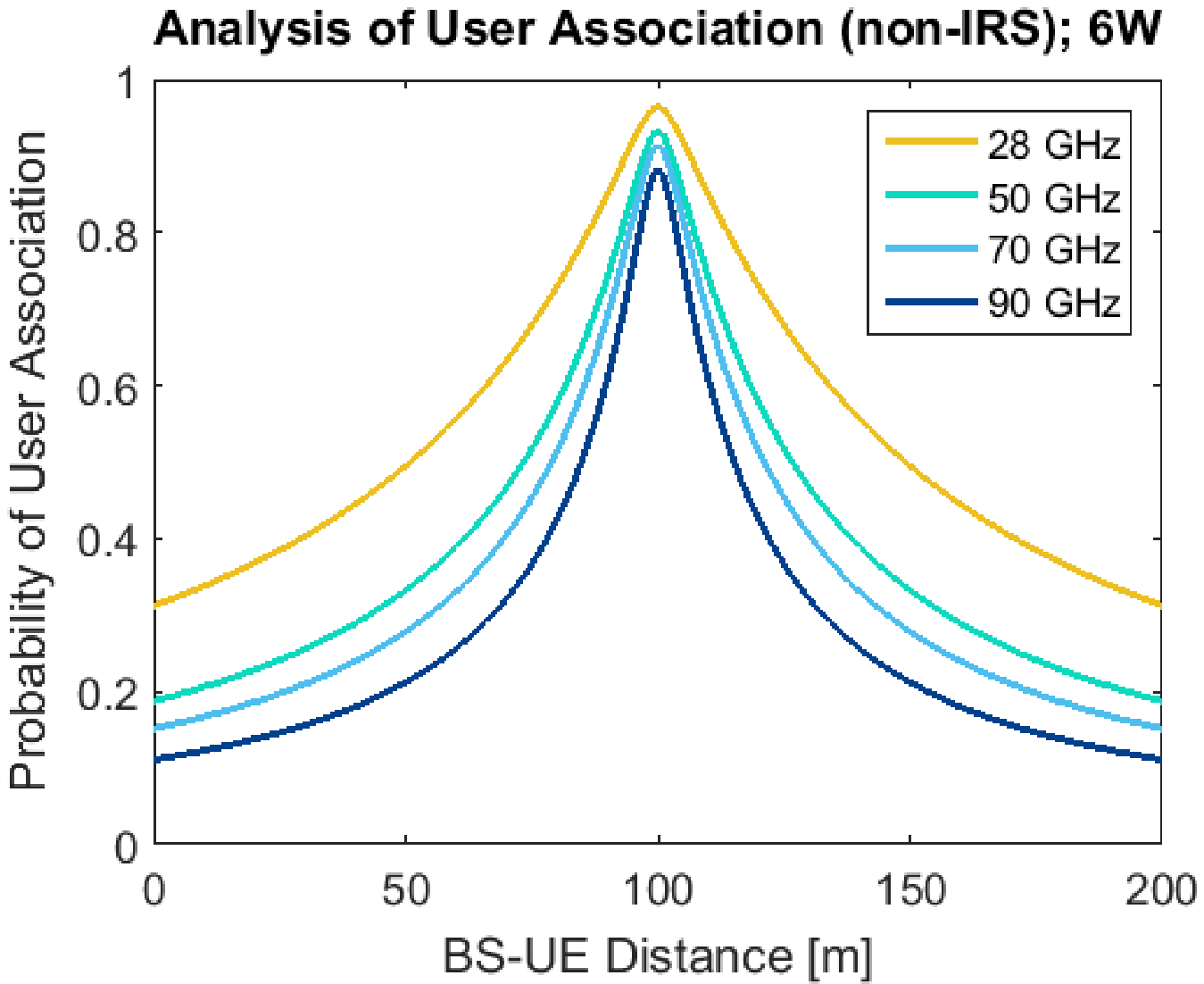}}
\vspace{3pt}
\centerline{\footnotesize{(b)}}
\caption{(a) Probability of user association (4 W), (b) Probability of user association (6 W).}
\label{fig}
\end{figure}

Fig. 3 (a)-(d) represent the probability of user association corresponding to the 28, 50, 70, and 90 GHz mmWave carriers, respectively for 6 W of transmit power of the conventional (non-IRS) micro base station.

\begin{figure}[htbp]
\centerline{\includegraphics[height=7.5cm, width=10.5cm]{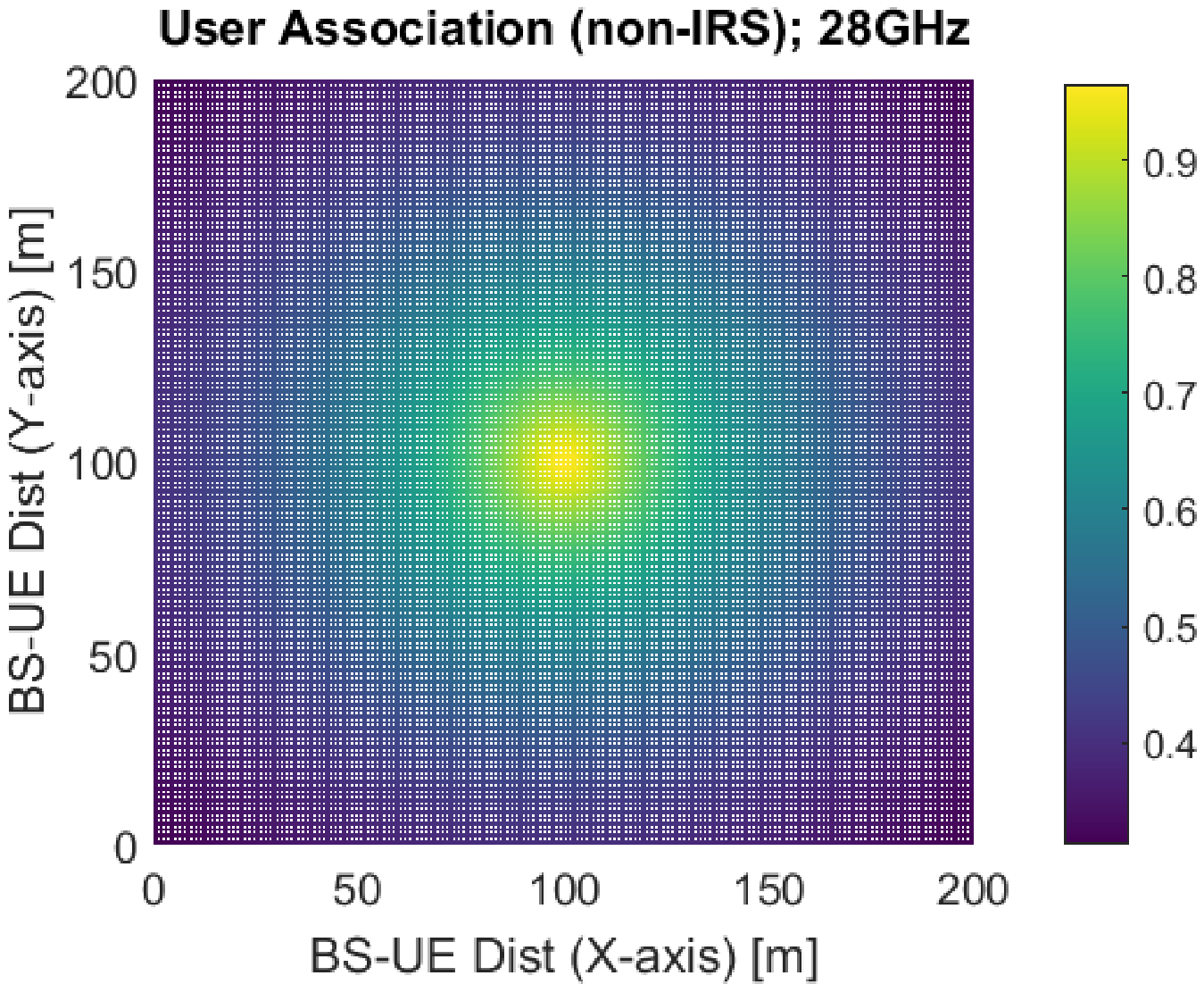}}
\vspace{3pt}
\centerline{\footnotesize{(a)}}
\label{fig}
\end{figure}

\begin{figure}[htbp]
\centerline{\includegraphics[height=7.5cm, width=10.5cm]{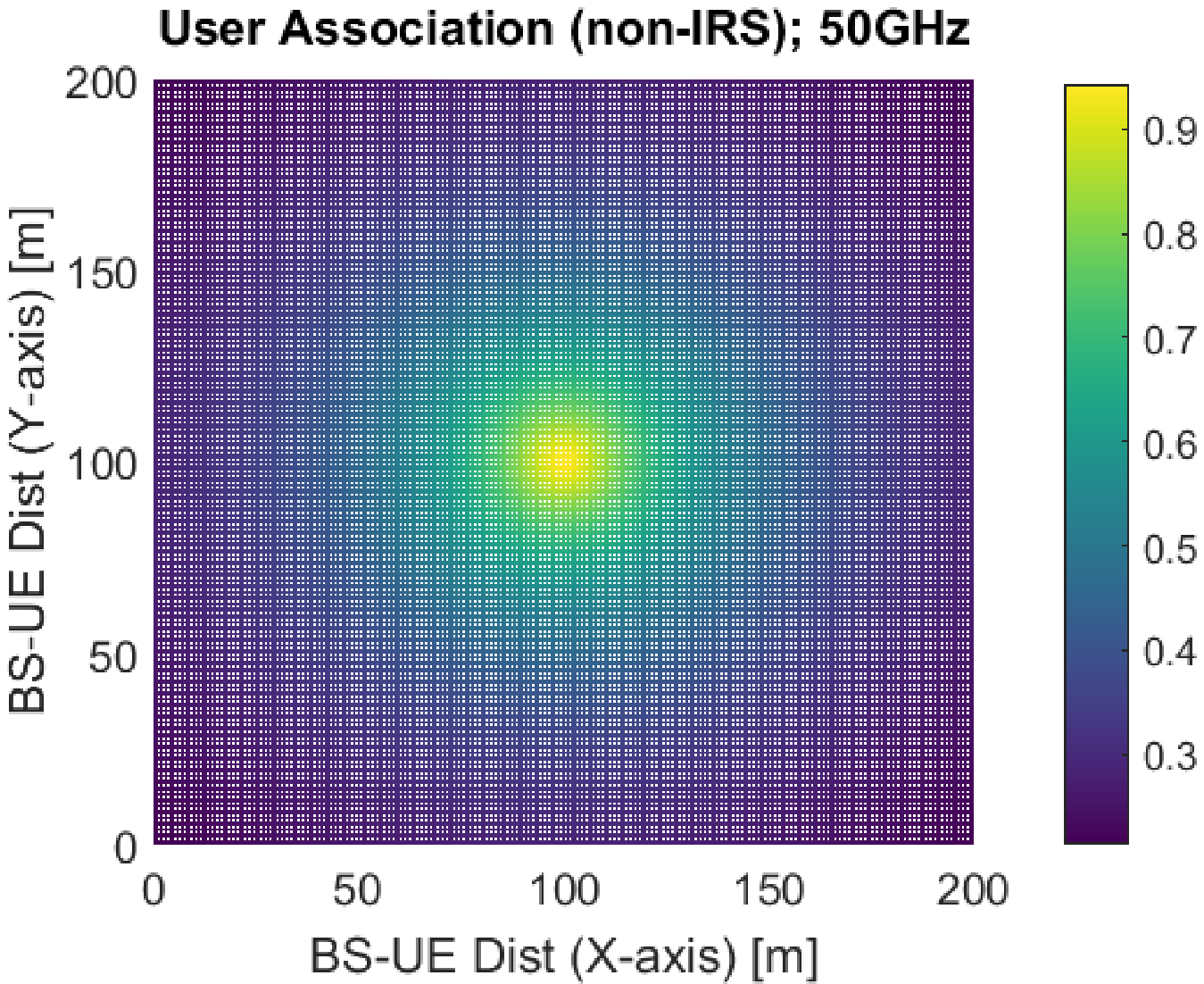}}
\vspace{3pt}
\centerline{\footnotesize{(b)}}
\label{fig}
\end{figure}

\begin{figure}[htbp]
\centerline{\includegraphics[height=7.5cm, width=10.5cm]{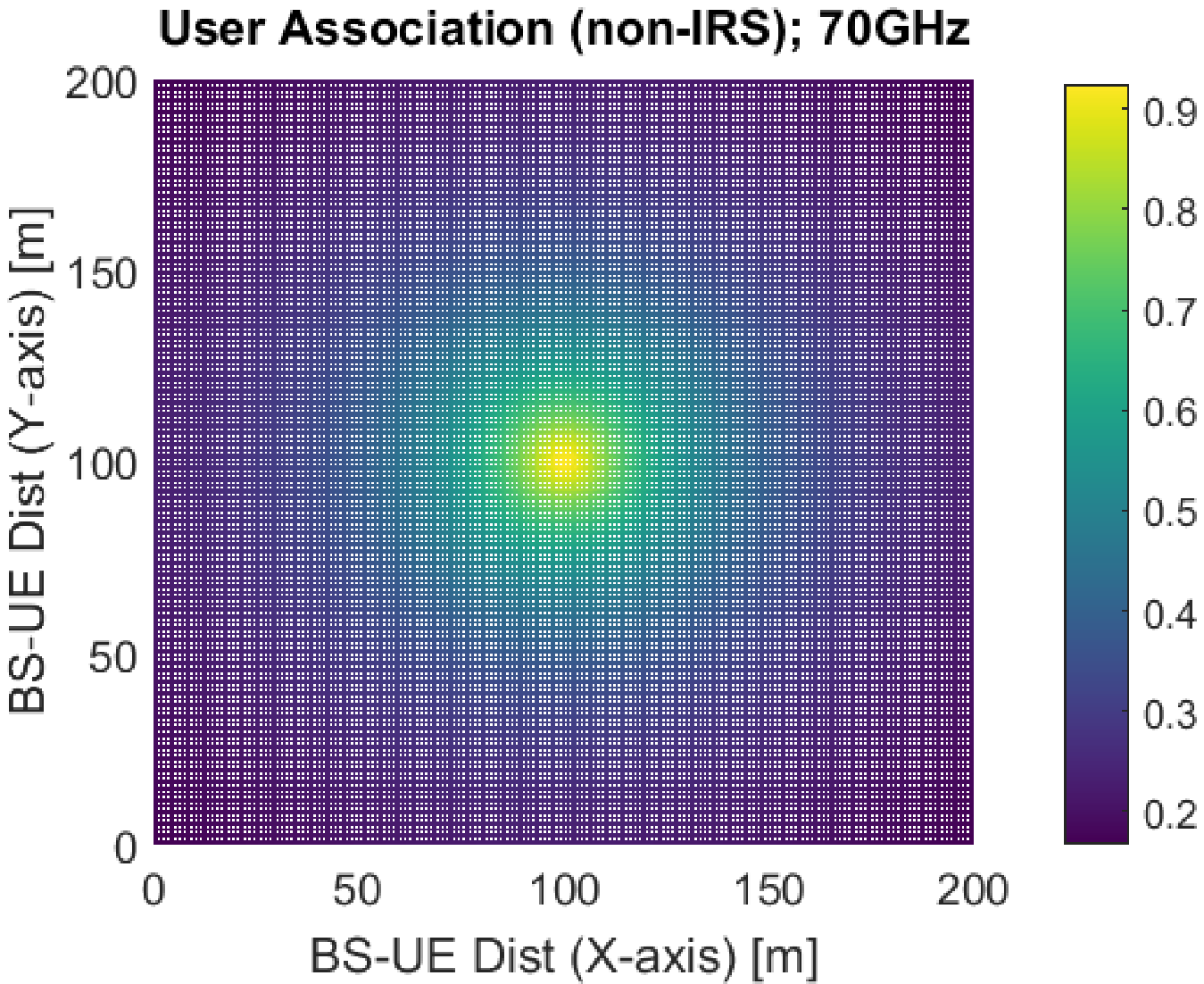}}
\vspace{3pt}
\centerline{\footnotesize{(c)}}
\label{fig}
\end{figure}

\begin{figure}[htbp]
\centerline{\includegraphics[height=7.5cm, width=10.5cm]{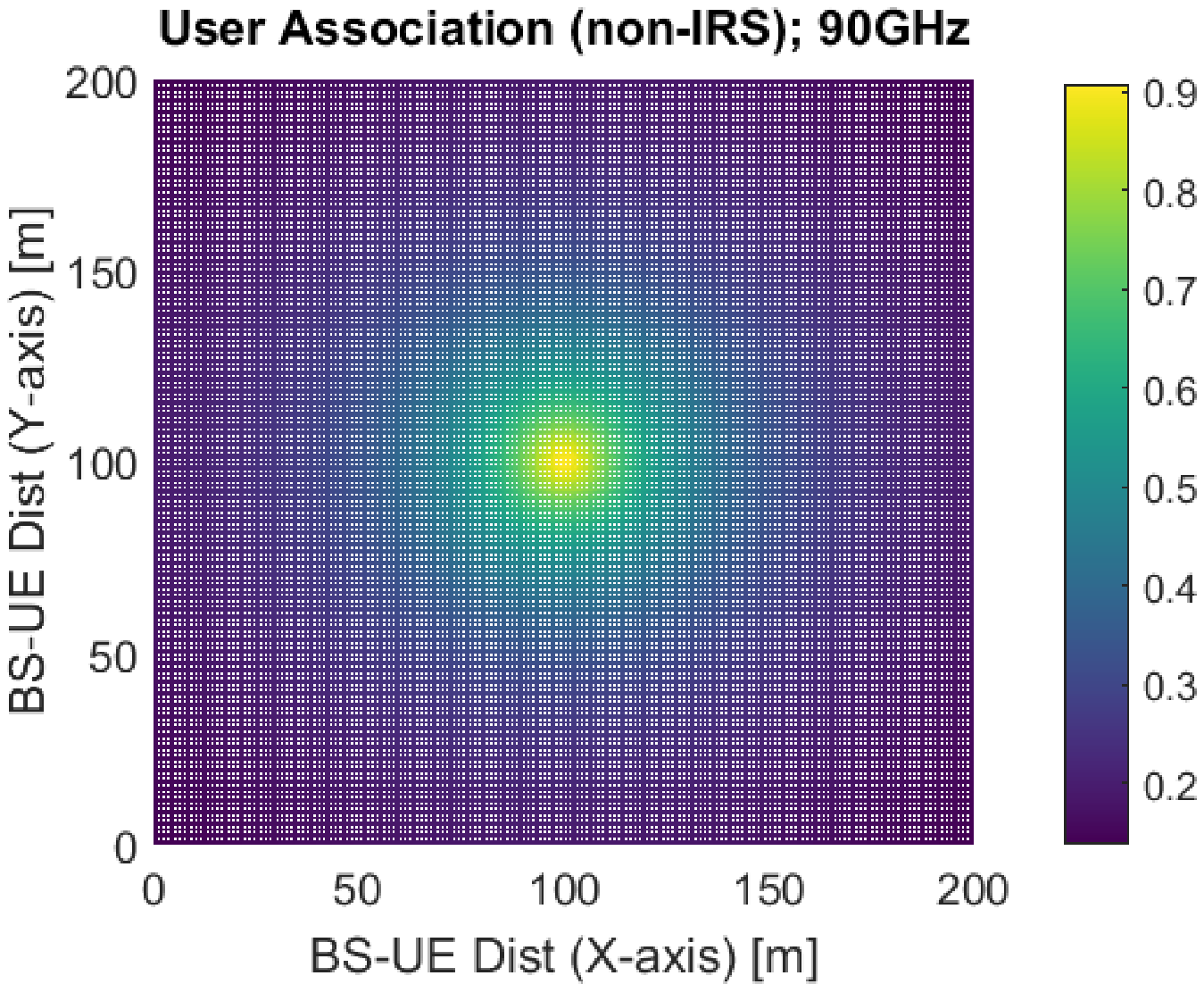}}
\vspace{3pt}
\centerline{\footnotesize{(d)}}
\caption{(a) Probability of user association (28 GHz), (b) Probability of user association (50 GHz), (c) Probability of user association (70 GHz), (d) Probability of user association (90 GHz).}
\label{fig}
\end{figure}

According to the observation of Figs. 2 and 3 it is comprehensible that, for 4 W of transmission power of micro base station (non-IRS), 28 GHz carrier achieves a maximum of 0.96 and a minimum of 0.28 probability of user association, 50 GHz carrier achieves a maximum of 0.92 and a minimum of 0.16 probability of user association, 70 GHz carrier achieves a maximum of 0.90 and a minimum of 0.13 probability of user association, and 90 GHz carrier achieves a maximum of 0.86 and a minimum of 0.09 probability of user association. An important clarification is that the maximum and minimum values are representing the probability of user association at the cell center and cell edge, respectively.

In the case of 6 W of transmission power of micro base station, 28 GHz carrier achieves a maximum of 0.97 and a minimum of 0.31 probability of user association, 50 GHz carrier achieves a maximum of 0.93 and a minimum of 0.19 probability of user association, 70 GHz carrier achieves a maximum of 0.92 and a minimum of 0.15 probability of user association, and 90 GHz carrier achieves a maximum of 0.88 and a minimum of 0.11 probability of user association. It can be said that the enhancement of the transmission power increases the probability of user association a bit but it is not much significant or sufficient for the cell-edge users.

Fig. 4 (a) shows the probability of user association for 1 W of transmit power of the IRS-assisted micro base station, 128 transmitter-receiver elements of the IRS, and 45$\degree$ of transmit-receive angles. Fig. 4 (b) shows the probability of user association in an IRS-assisted micro base station for 2 W of transmit power, 128 transmitter-receiver elements, and 45$\degree$ of transmit-receive angles. Fig. 4 (c) shows the probability of user association in an IRS-assisted micro base station for 2 W of transmit power, 128 transmitter-receiver elements, and 60$\degree$ of transmit-receive angles. Fig. 4 (d) shows the probability of user association in an IRS-assisted micro base station for 2 W of transmit power, 256 transmitter-receiver elements, and 45$\degree$ of transmit-receive angles.

\begin{figure}[htbp]
\centerline{\includegraphics[height=7.5cm, width=10.5cm]{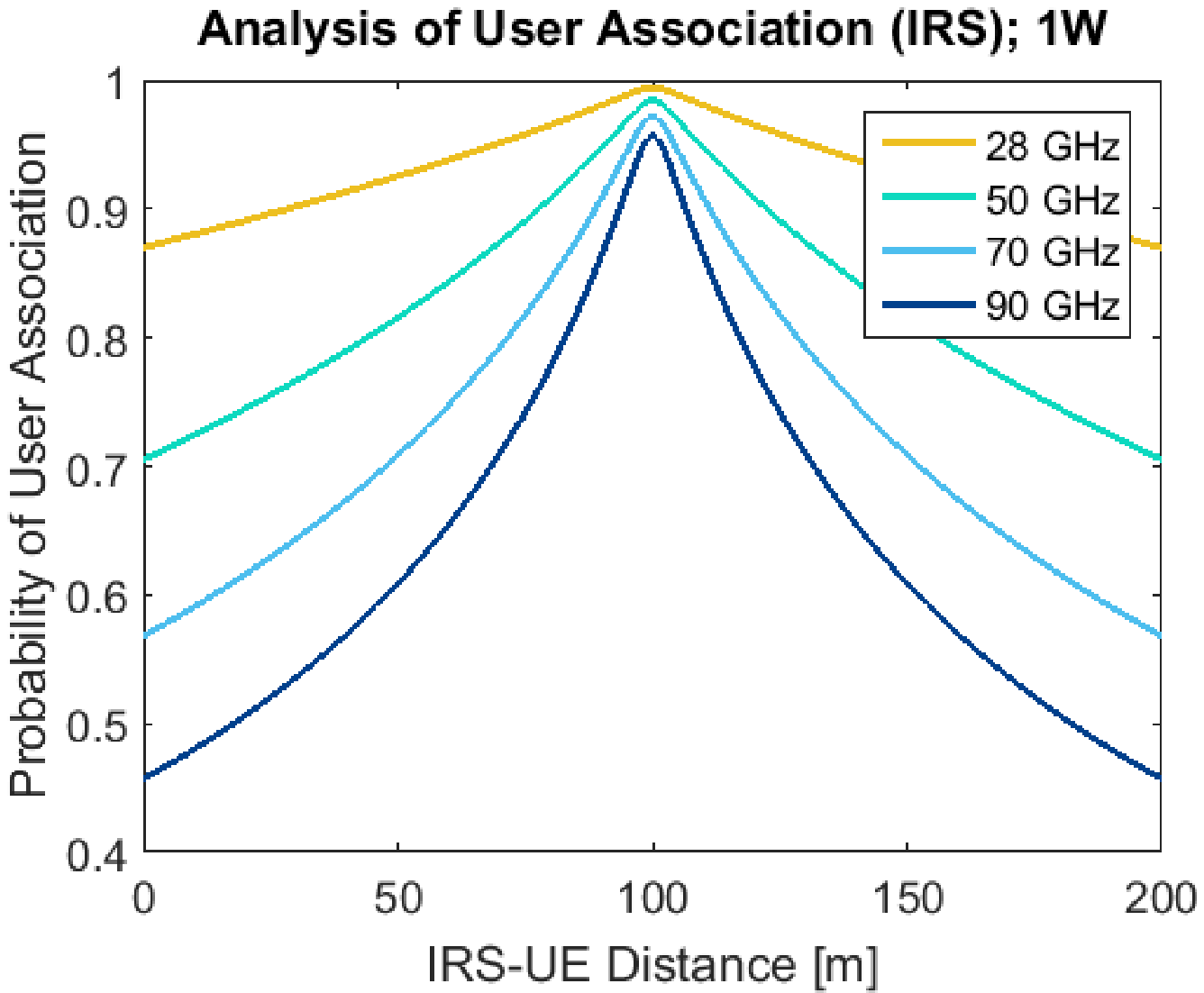}}
\vspace{3pt}
\centerline{\footnotesize{(a)}}
\label{fig}
\end{figure}

\begin{figure}[htbp]
\centerline{\includegraphics[height=7.5cm, width=10.5cm]{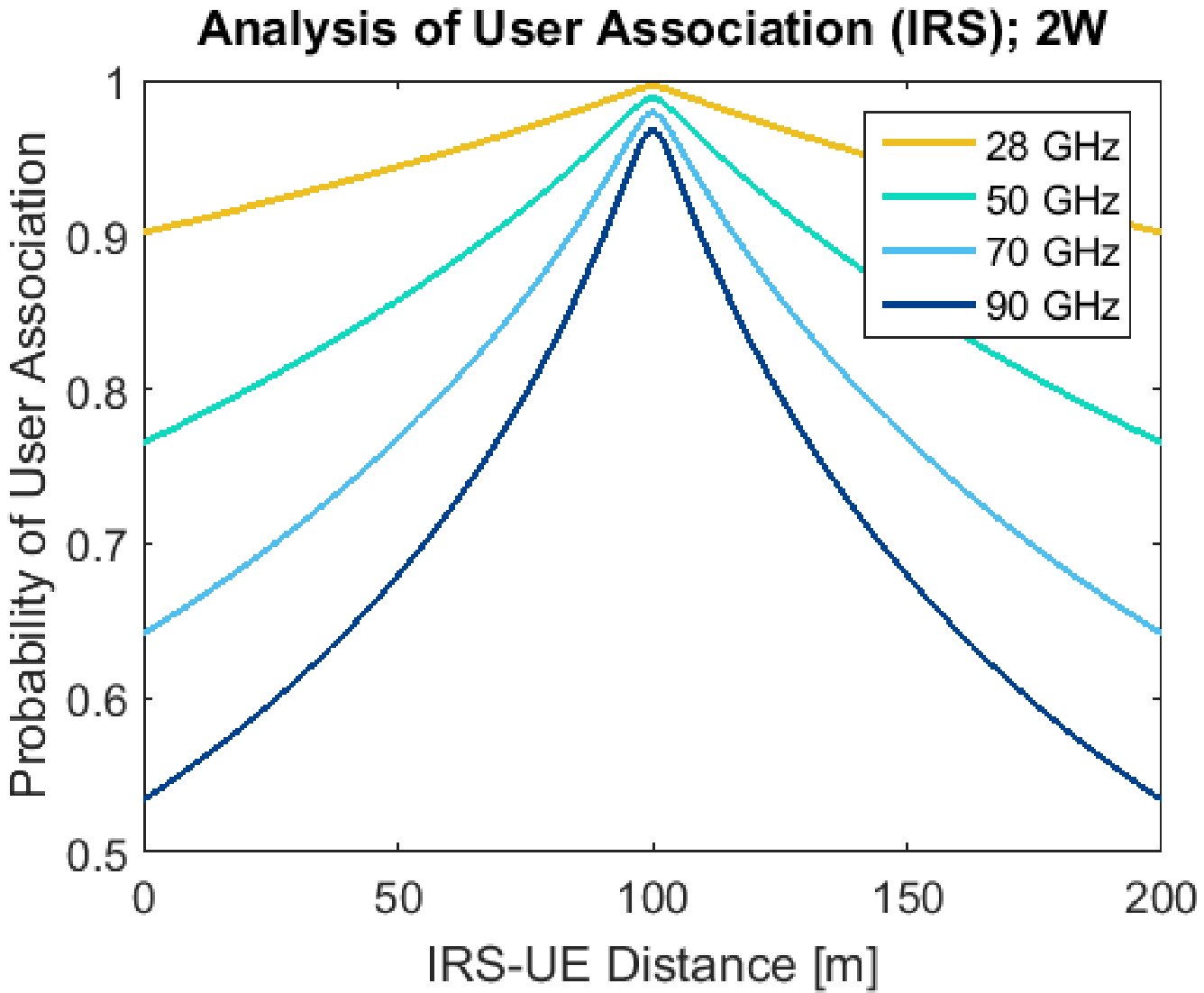}}
\vspace{3pt}
\centerline{\footnotesize{(b)}}
\label{fig}
\end{figure}

\begin{figure}[htbp]
\centerline{\includegraphics[height=7.5cm, width=10.5cm]{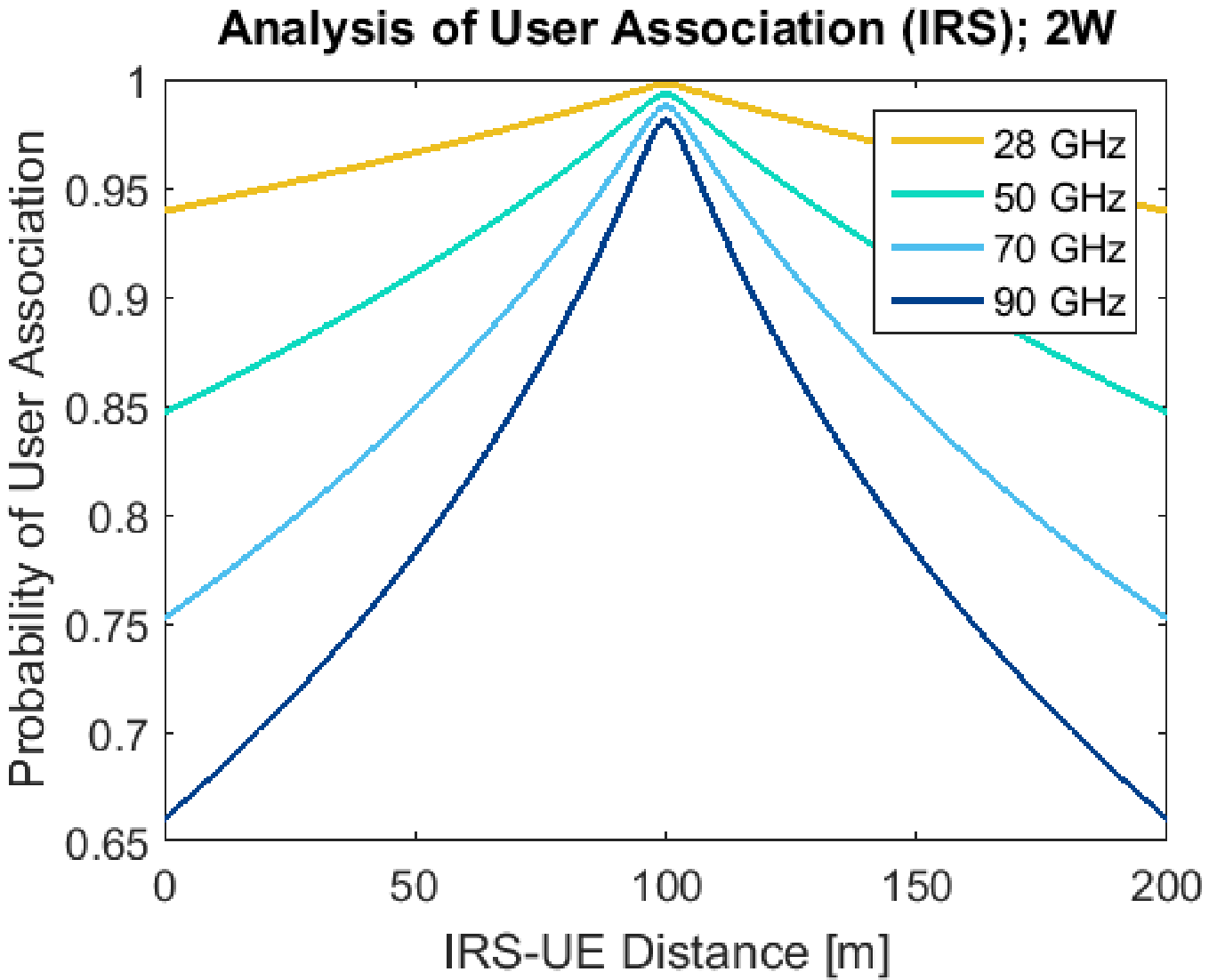}}
\vspace{3pt}
\centerline{\footnotesize{(c)}}
\label{fig}
\end{figure}

\begin{figure}[htbp]
\centerline{\includegraphics[height=7.5cm, width=10.5cm]{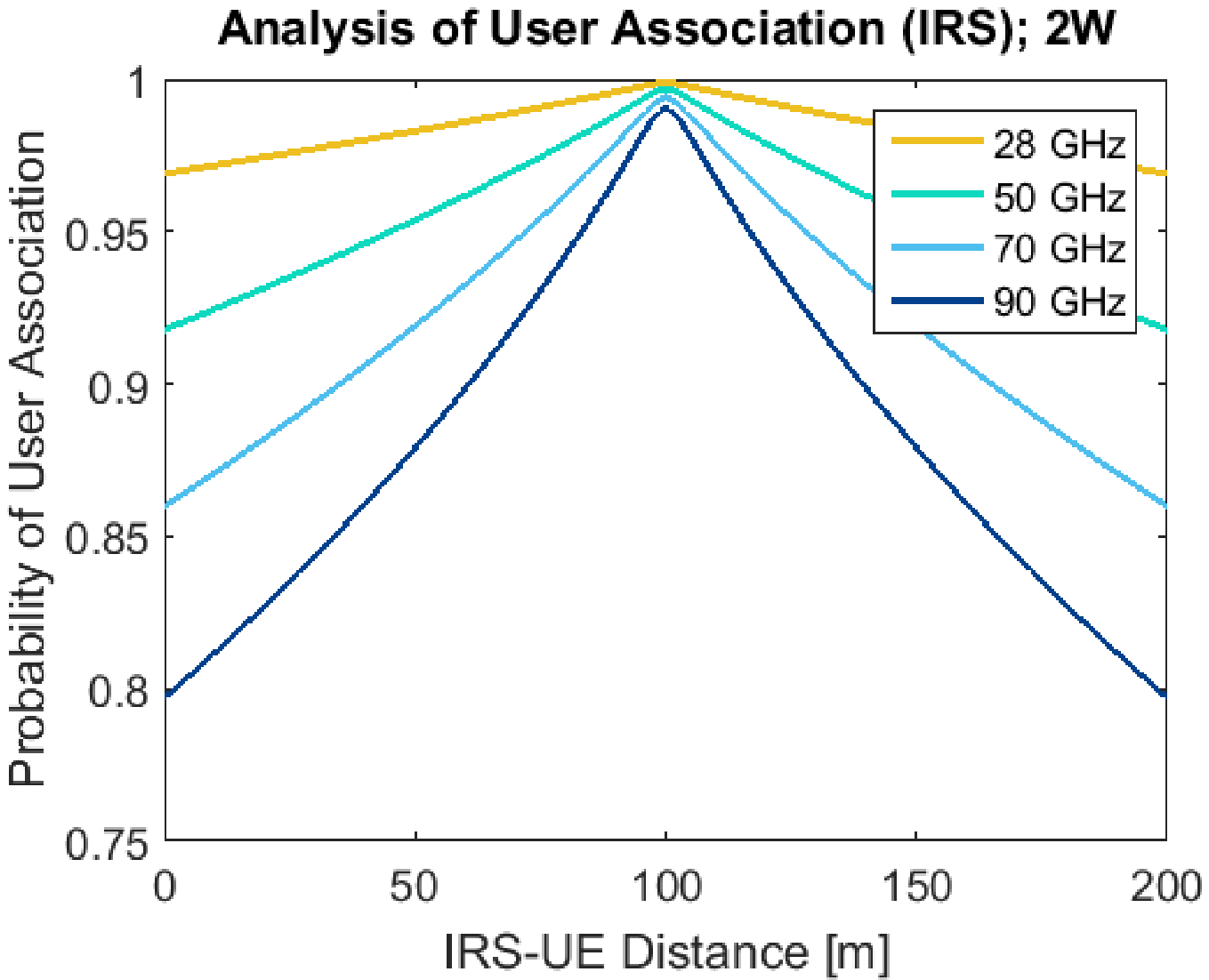}}
\vspace{3pt}
\centerline{\footnotesize{(d)}}
\caption{(a) Probability of user association (1 W, 128 transmitter and receiver elements, 45$\degree$ transmit and receive angles), (b) Probability of user association (2 W, 128 transmitter and receiver elements, 45$\degree$ transmit and receive angles), (c) Probability of user association (2 W, 128 transmitter and receiver elements, 60$\degree$ transmit and receive angles), (d) Probability of user association (2 W, 256 transmitter and receiver elements, 45$\degree$ transmit and receive angles).}
\label{fig}
\end{figure}

Fig. 5 (a) visualizes the probability of user association in the context of an IRS-enhanced micro cell for 1 W of transmit power of the base station, 128 transmitter-receiver elements of the IRS, 45$\degree$ of transmit-receive angles, and a 28 GHz carrier. Fig. 5 (b) visualizes the IRS-enhanced micro base station’s probability of user association for 1 W of transmit power, 128 transmitter-receiver elements, 45$\degree$ of transmit-receive angles, and 90 GHz carrier. Fig. 5 (c) visualizes the probability of user association in the case of an IRS-enhanced micro base station for 2 W of transmit power, 45$\degree$ of transmit-receive angles, 128 transmitter-receiver elements of the IRS, and a 90 GHz carrier. Fig. 5 (d) visualizes the probability of user association in the case of an IRS-enhanced micro base station for 2 W of transmit power, 256 transmitter-receiver elements, 45$\degree$ of transmit-receive angles, and a 90 GHz carrier.

\begin{figure}[htbp]
\centerline{\includegraphics[height=7.5cm, width=10.5cm]{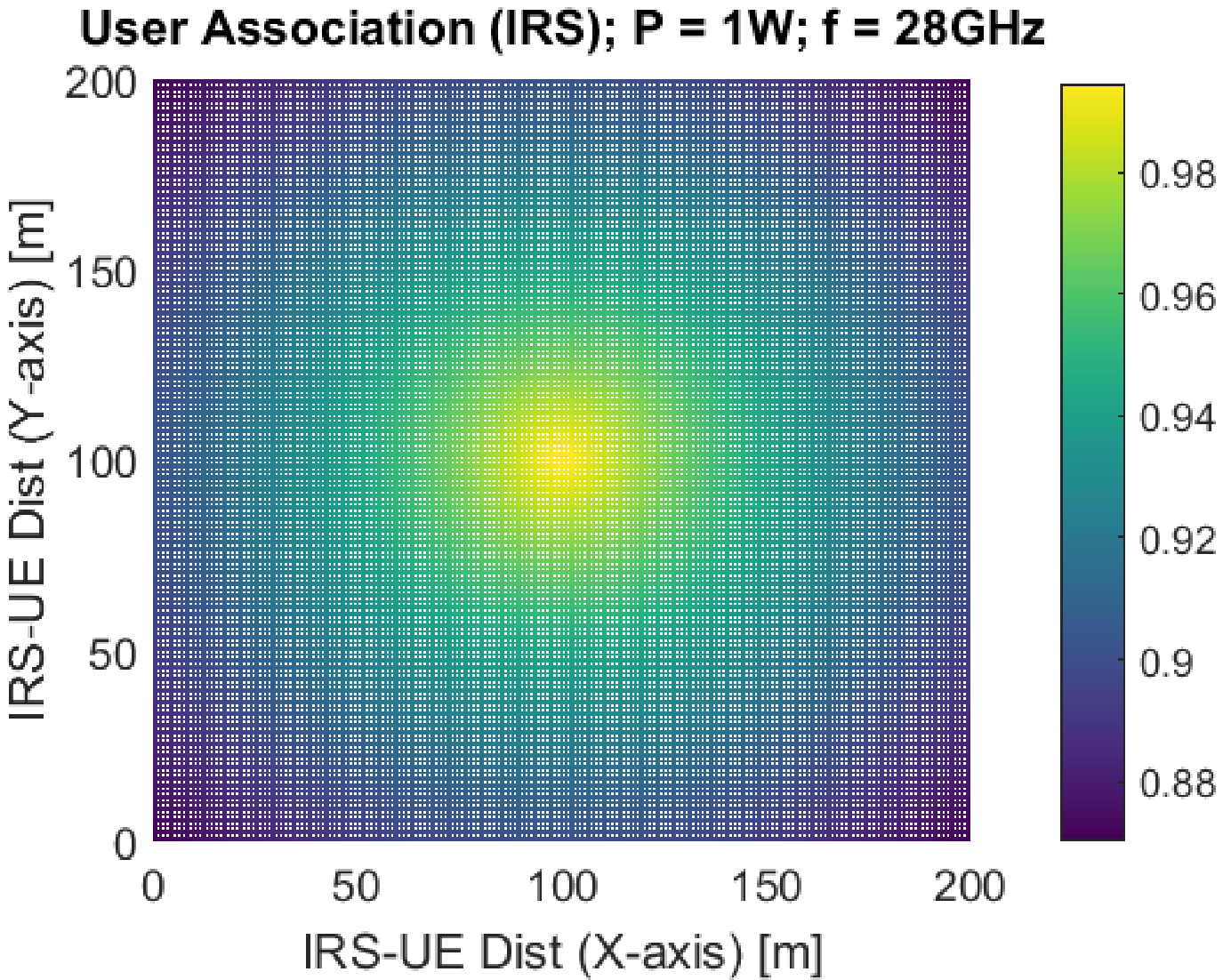}}
\vspace{3pt}
\centerline{\footnotesize{(a)}}
\label{fig}
\end{figure}

\begin{figure}[htbp]
\centerline{\includegraphics[height=7.5cm, width=10.5cm]{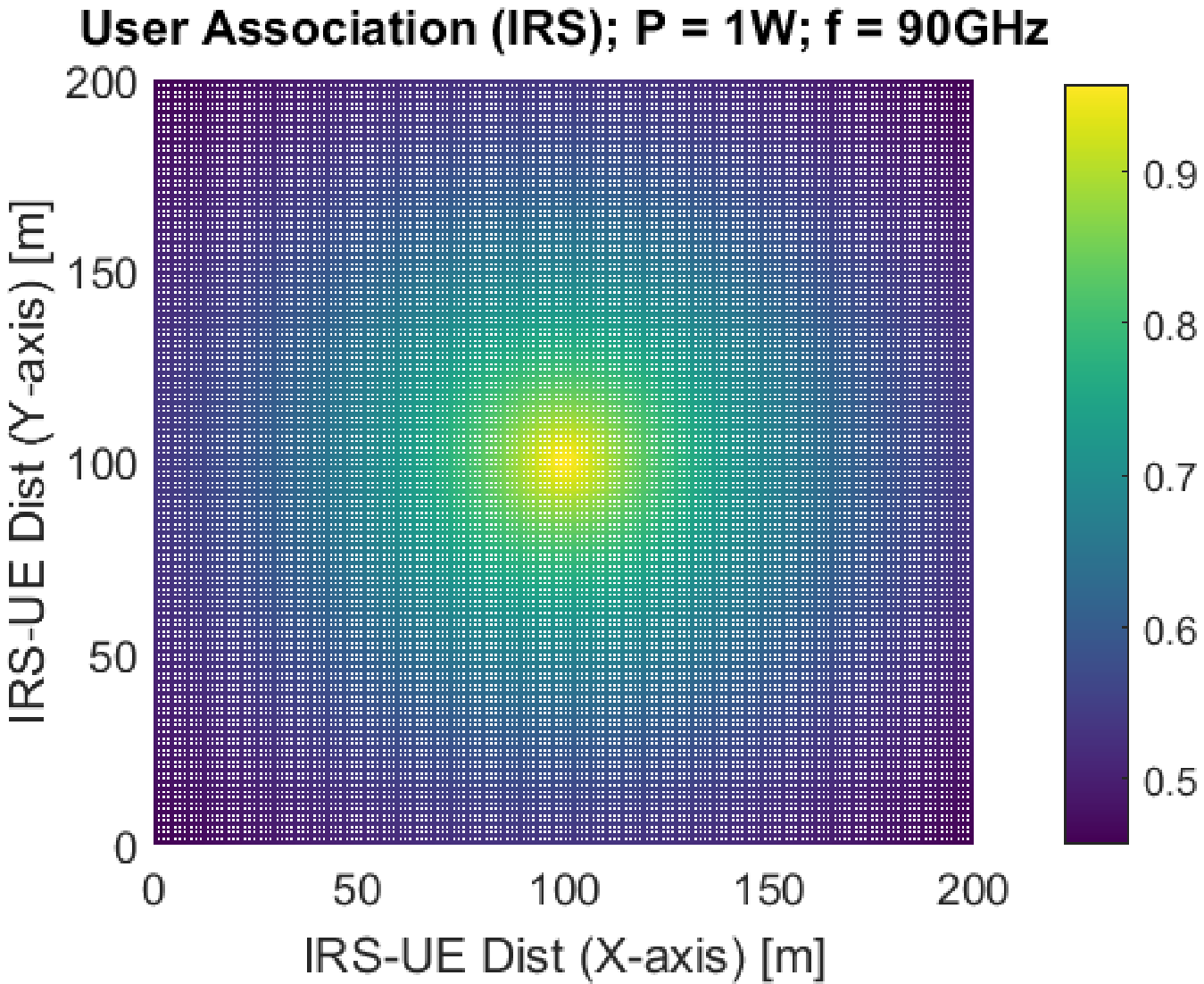}}
\vspace{3pt}
\centerline{\footnotesize{(b)}}
\label{fig}
\end{figure}

\begin{figure}[htbp]
\centerline{\includegraphics[height=7.5cm, width=10.5cm]{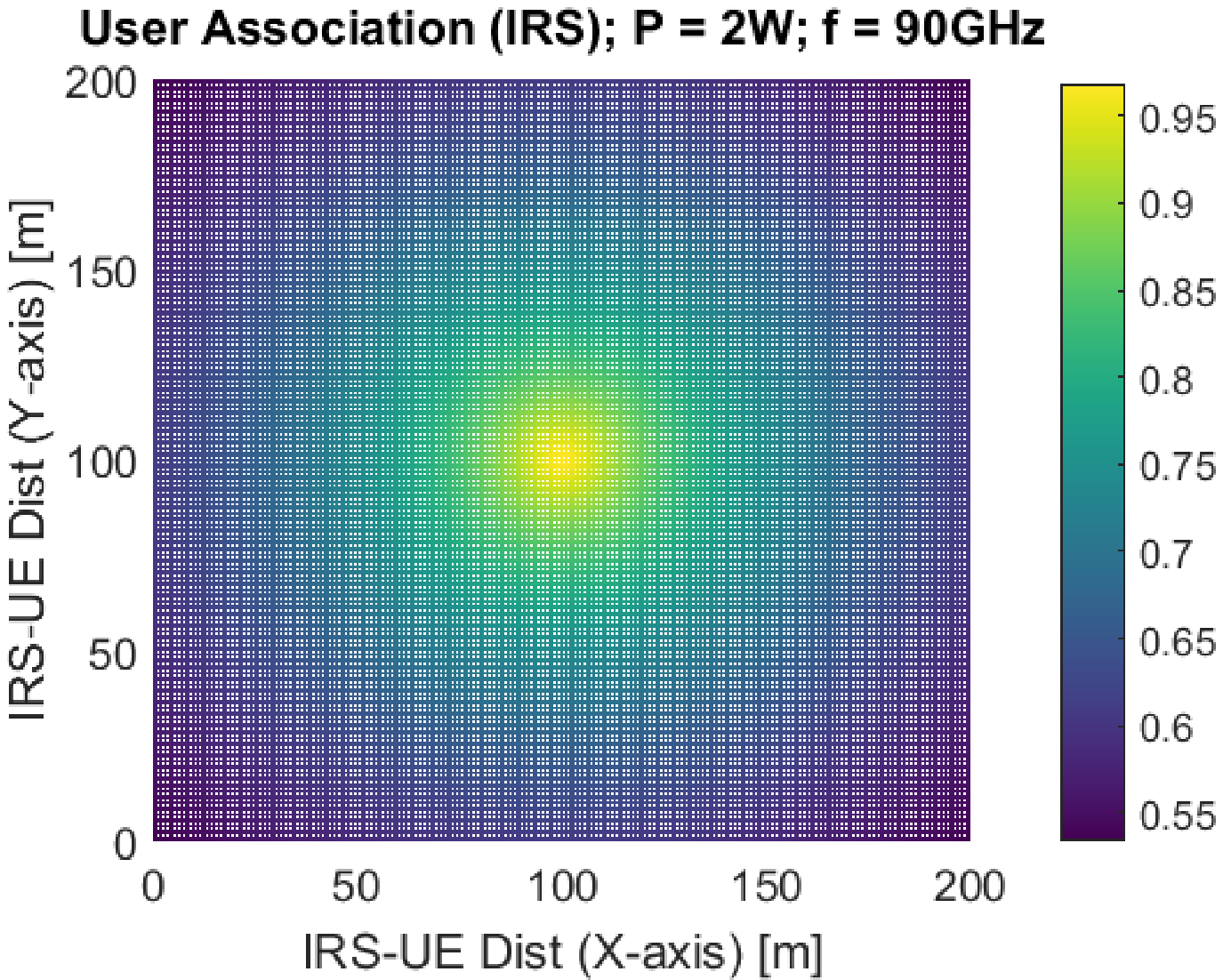}}
\vspace{3pt}
\centerline{\footnotesize{(c)}}
\label{fig}
\end{figure}

\begin{figure}[htbp]
\centerline{\includegraphics[height=7.5cm, width=10.5cm]{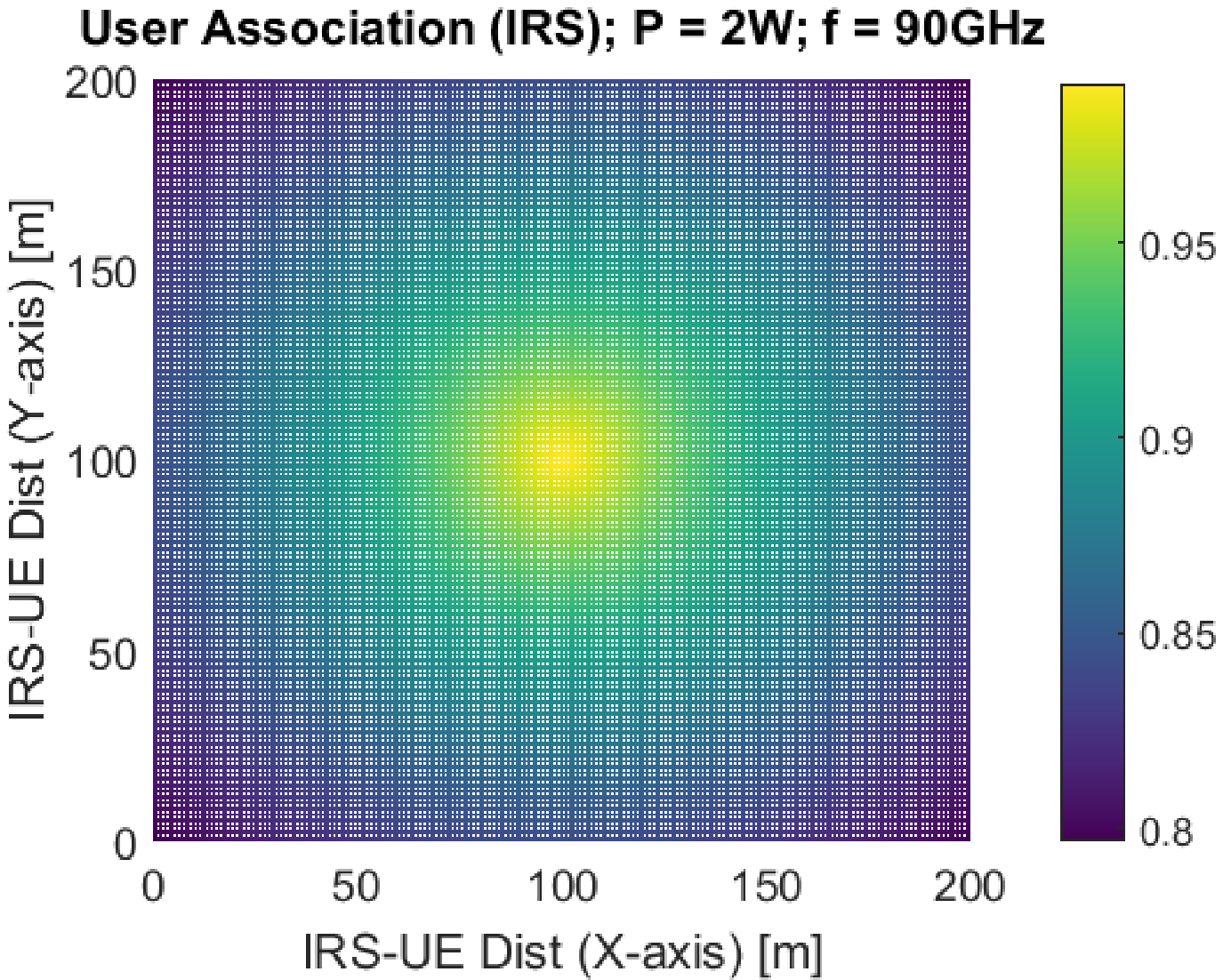}}
\vspace{3pt}
\centerline{\footnotesize{(d)}}
\caption{(a) Probability of user association (1 W, 128 transmitter and receiver elements, 28 GHz), (b) Probability of user association (1 W, 128 transmitter and receiver elements, 90 GHz), (c) Probability of user association (2 W, 128 transmitter and receiver elements, 90 GHz), (d) Probability of user association (2W, 256 transmitter and receiver elements, 90 GHz).}
\label{fig}
\end{figure}

Observing Fig. 4 (a)-(d) it is realizable that, for 1 W of transmission power of IRS-assisted micro base station, 128 transmitter-receiver elements of IRS, and 45$\degree$ transmit (base station-to-IRS)-receive (IRS-to-user) angles the maximum and minimum values of the probability of user association for 28, 50, 70, and 90 GHz carriers are 0.9945 and 0.87, 0.9846 and 0.70, 0.9724 and 0.57, and 0.9575 and 0.45, respectively.

In the case of 2 W transmission power, 128 transmitter-receiver elements of IRS, and 45$\degree$ transmit-receive angles the maximum and minimum values of the probability of user association for 28, 50, 70, and 90 GHz carriers are 0.9959 and 0.90, 0.9887 and 0.77, 0.9796 and 0.64, and 0.9684 and 0.53, respectively.

In the case of 2 W transmission power, 128 transmitter-receiver elements of IRS, and 60$\degree$ transmit-receive angles the maximum and minimum values of the probability of user association for 28, 50, 70, and 90 GHz carriers are 0.9976 and 0.94, 0.9933 and 0.85, 0.9879 and 0.75, and 0.9812 and 0.66, respectively.

In the case of 2 W transmission power, 256 transmitter-receiver elements of IRS, and 45$\degree$ transmit-receive angles the maximum and minimum values of the probability of user association for 28, 50, 70, and 90 GHz carriers are 0.9988 and 0.97, 0.9967 and 0.92, 0.994 and 0.86, and 0.9906 and 0.80, respectively.

Through the observation, it can be stated that, with a lower level of transmission power (i.e., 1 W), a lower number of transmitter -receiver elements (128 elements) the lower level mmWave carrier such as 28 GHz can provide sufficient coverage to the entire coverage region. For higher-level mmWave carriers such as 90 GHz a bit high transmission power (i.e., 2 W), an increased number of transmitter-receiver elements (i.e., 256 elements) are required to ensure sufficient coverage to the entire coverage region, especially for the cell-edge users. The performance of the network depends upon transmit-receive angles as well (Fig. 4(c)). The research obtained that, by maintaining an efficient trade-off among network resources namely carrier, transmit power, number of transmitter-receiver elements, etc. a significantly enhanced coverage can be achieved in an IRS-enhanced micro cell compared to the conventional micro cell. The cell-edge users are as well achieving significant association probability through the deployment of IRS.

Comparing the measurement results of both the conventional non-IRS and IRS-assisted micro base stations the work figures out that the deployment of IRS maximizes or enhances the probability of user association significantly with a reduced transmit power of the micro base station. For example, in the case of a conventional non-IRS micro base station, for 6 W of transmission power of micro base station, 28 GHz mmWave carrier achieves a maximum of 0.97 and a minimum of 0.31 probability of user association. Whereas, in the case of an IRS-assisted micro base station with 1 W of transmission power, 128 transmitter-receiver elements, and 45$\degree$ transmit (base station-to-IRS)-receive (IRS-to-user) angles the maximum and minimum values of the probability of user association for the 28 GHz carrier are 0.9945 and 0.87. If the transmit power is increased to 2 W (128 transmitter-receiver elements and 45$\degree$ transmit-receive angles) the maximum and minimum values of the probability of user association for the 28 GHz carrier are 0.9959 and 0.90.

The deployment of IRS in a micro cell reduces the energy consumption up to 67\% - 83\% by reducing the transmit power of the micro base station from 6 W (conventional micro cell [125]) to 1-2 W (IRS-assisted micro cell). Due to the placement of IRS with such a reduced power the cell-edge users achieve significantly favorable coverage when an efficient resource orchestration is confirmed and this is better realizable through Fig. 5(a)-(d).

Another notable finding of the research is that for the mm Wave-based transmission utilizing IRS is one of the most favorable solutions to enhance the entire network capacity.

Notable information is that the work during the literature review found a total of thirty-five articles in scholarly research databases such as IEEE Xplore, Science Direct, and Springer Link relative to the analysis of user association in 6G networks. The review found four articles on IRS/RIS-assisted user association analysis for mmWave networks and only two specific to 6G.
}
\vspace{18pt}

\RaggedRight{\textbf{\Large 5.\hspace{10pt} Conclusion}}\\
\vspace{18pt}
\justifying \noindent {The work aimed to maximize the probability of user association of a micro cell of a two-tier network assuring energy efficiency. Therefore, the research formulated a measurement model including required equations to measure and compares the probability of user association in the context of a conventional micro cell and IRS-assisted micro cell of a two-tier network in which the micro base stations are operating under macro cell. The work derives that the deployment of IRS significantly maximizes the performance of the micro base stations or cells assuring significant energy efficiency. Moreover, the IRS notably enhances the probability of user association for cell-edge users as well when an efficient orchestration of network resources is assured.}
\vspace{18pt}

\RaggedRight{\textbf{\Large References}}\\
\vspace{12pt}

\justifying{
1.	Mohamed I. AlHajri, Nazar T. Ali, and Raed M. Shubair. "Classification of indoor environments for IoT applications: A machine learning approach." IEEE Antennas and Wireless Propagation Letters 17, no. 12 (2018): 2164-2168.

2.	M. I. AlHajri, A. Goian, M. Darweesh, R. AlMemari, R. M. Shubair, L. Weruaga, and A. R. Kulaib. "Hybrid RSS-DOA technique for enhanced WSN localization in a correlated environment." In 2015 International Conference on Information and Communication Technology Research (ICTRC), pp. 238-241. IEEE, 2015.

3.	Mohamed I. AlHajri, Nazar T. Ali, and Raed M. Shubair. "Indoor localization for IoT using adaptive feature selection: A cascaded machine learning approach." IEEE Antennas and Wireless Propagation Letters 18, no. 11 (2019): 2306-2310.

4.	M. A. Al-Nuaimi, R. M. Shubair, and K. O. Al-Midfa. "Direction of arrival estimation in wireless mobile communications using minimum variance distortionless response." In The Second International Conference on Innovations in Information Technology (IIT’05), pp. 1-5. 2005.

5.	Fahad Belhoul, Raed M. Shubair, and Mohammed E. Al-Mualla. "Modelling and performance analysis of DOA estimation in adaptive signal processing arrays." In ICECS, pp. 340-343. 2003.

6.	Ali Hakam, Raed M. Shubair, and Ehab Salahat. "Enhanced DOA estimation algorithms using MVDR and MUSIC." In 2013 International Conference on Current Trends in Information Technology (CTIT), pp. 172-176. IEEE, 2013.

7.	Ebrahim M. Al-Ardi, Raed M. Shubair, and Mohammed E. Al-Mualla. "Direction of arrival estimation in a multipath environment: An overview and a new contribution." Applied Computational Electromagnetics Society Journal 21, no. 3 (2006): 226.

8.	R. M. Shubair. "Robust adaptive beamforming using LMS algorithm with SMI initialization." In 2005 IEEE Antennas and Propagation Society International Symposium, vol. 4, pp. 2-5. IEEE, 2005.

9.	R. M. Shubair and A. Al-Merri. "Robust algorithms for direction finding and adaptive beamforming: performance and optimization." In The 2004 47th Midwest Symposium on Circuits and Systems, 2004. MWSCAS'04., vol. 2, pp. II-II. IEEE, 2004.

10.	Pradeep Kumar Singh, Bharat K. Bhargava, Marcin Paprzycki, Narottam Chand Kaushal, and Wei-Chiang Hong, eds. Handbook of wireless sensor networks: issues and challenges in current Scenario's. Vol. 1132. Berlin/Heidelberg, Germany: Springer, 2020.

11.	R. M. Shubair and Y. L. Chow. "A closed-form solution of vertical dipole antennas above a dielectric half-space." IEEE transactions on antennas and propagation 41, no. 12 (1993): 1737-1741.

12.	Ebrahim M. Al-Ardi, Raed M. Shubair, and Mohammed E. Al-Mualla. "Computationally efficient DOA estimation in a multipath environment using covariance differencing and iterative spatial smoothing." In 2005 IEEE International Symposium on Circuits and Systems, pp. 3805-3808. IEEE, 2005.

13.	R. M. Shubair and W. Jessmi. "Performance analysis of SMI adaptive beamforming arrays for smart antenna systems." In 2005 IEEE Antennas and Propagation Society International Symposium, vol. 1, pp. 311-314. IEEE, 2005.

14.	E. M. Al-Ardi, R. M. Shubair, and M. E. Al-Mualla. "Investigation of high-resolution DOA estimation algorithms for optimal performance of smart antenna systems." (2003): 460-464.

15.	E. M. Al-Ardi, Raed M. Shubair, and M. E. Al-Mualla. "Performance evaluation of direction finding algorithms for adapative antenna arrays." In 10th IEEE International Conference on Electronics, Circuits and Systems, 2003. ICECS 2003. Proceedings of the 2003, vol. 2, pp. 735-738. IEEE, 2003.

16.	M. I. AlHajri, N. Alsindi, N. T. Ali, and R. M. Shubair. "Classification of indoor environments based on spatial correlation of RF channel fingerprints." In 2016 IEEE international symposium on antennas and propagation (APSURSI), pp. 1447-1448. IEEE, 2016.

17.	Mohamed AlHajri, Abdulrahman Goian, Muna Darweesh, Rashid AlMemari, Raed Shubair, Luis Weruaga, and Ahmed AlTunaiji. "Accurate and robust localization techniques for wireless sensor networks." arXiv preprint arXiv:1806.05765 (2018).

18.	Goian, Mohamed I. AlHajri, Raed M. Shubair, Luis Weruaga, Ahmed Rashed Kulaib, R. AlMemari, and Muna Darweesh. "Fast detection of coherent signals using pre-conditioned root-MUSIC based on Toeplitz matrix reconstruction." In 2015 IEEE 11th International Conference on Wireless and Mobile Computing, Networking and Communications (WiMob), pp. 168-174. IEEE, 2015.

19.	Raed M. Shubair, and Ali Hakam. "Adaptive beamforming using variable step-size LMS algorithm with novel ULA array configuration." In 2013 15th IEEE International Conference on Communication Technology, pp. 650-654. IEEE, 2013.

20.	R. M. Shubair, and A. Merri. "Convergence of adaptive beamforming algorithms for wireless communications." In Proc. IEEE and IFIP International Conference on Wireless and Optical Communications Networks, pp. 6-8. 2005.

21.	Zhenghua Chen, Mohamed I. AlHajri, Min Wu, Nazar T. Ali, and Raed M. Shubair. "A novel real-time deep learning approach for indoor localization based on RF environment identification." IEEE Sensors Letters 4, no. 6 (2020): 1-4.

22.	R. M. Shubair, A. Merri, and W. Jessmi. "Improved adaptive beamforming using a hybrid LMS/SMI approach." In Second IFIP International Conference on Wireless and Optical Communications Networks, 2005. WOCN 2005., pp. 603-606. IEEE, 2005.

23.	Mohamed I. AlHajri, Nazar T. Ali, and Raed M. Shubair. "A machine learning approach for the classification of indoor environments using RF signatures." In 2018 IEEE Global Conference on Signal and Information Processing (GlobalSIP), pp. 1060-1062. IEEE, 2018.

24.	E. M. Al-Ardi, R. M. Shubair, and M. E. Al-Mualla. "Performance evaluation of the LMS adaptive beamforming algorithm used in smart antenna systems." In 2003 46th Midwest Symposium on Circuits and Systems, vol. 1, pp. 432-435. IEEE, 2003.

25.	Satish R. Jondhale, Raed Shubair, Rekha P. Labade, Jaime Lloret, and Pramod R. Gunjal. "Application of supervised learning approach for target localization in wireless sensor network." In Handbook of Wireless Sensor Networks: Issues and Challenges in Current Scenario's, pp. 493-519. Springer, Cham, 2020.

26.	Raed Shubair, and Rashid Nuaimi. "Displaced sensor array for improved signal detection under grazing incidence conditions." Progress in Electromagnetics Research 79 (2008): 427-441.

27.	Raed M. Shubair, Abdulrahman S. Goian, Mohamed I. AlHajri, and Ahmed R. Kulaib. "A new technique for UCA-based DOA estimation of coherent signals." In 2016 16th Mediterranean Microwave Symposium (MMS), pp. 1-3. IEEE, 2016.

28.	M. I. AlHajri, R. M. Shubair, L. Weruaga, A. R. Kulaib, A. Goian, M. Darweesh, and R. AlMemari. "Hybrid method for enhanced detection of coherent signals using circular antenna arrays." In 2015 IEEE International Symposium on Antennas and Propagation \& USNC/URSI National Radio Science Meeting, pp. 1810-1811. IEEE, 2015.

29.	WafaNjima, Marwa Chafii, ArseniaChorti, Raed M. Shubair, and H. Vincent Poor. "Indoor localization using data augmentation via selective generative adversarial networks." IEEE Access 9 (2021): 98337-98347.

30.	Raed M. Shubair. "Improved smart antenna design using displaced sensor array configuration." Applied Computational Electromagnetics Society Journal 22, no. 1 (2007): 83.

31.	R. M. Shubair, and A. Merri. "A convergence study of adaptive beamforming algorithms used in smart antenna systems." In 11th International Symposium on Antenna Technology and Applied Electromagnetics [ANTEM 2005], pp. 1-5. IEEE, 2005.

32.	E. M. Ardi, , R. M. Shubair, and M. E. Mualla. "Adaptive beamforming arrays for smart antenna systems: A comprehensive performance study." In IEEE Antennas and Propagation Society Symposium, 2004., vol. 3, pp. 2651-2654. IEEE, 2004.

33.	Mohamed Ibrahim Alhajri, N. T. Ali, and R. M. Shubair. "2.4 ghz indoor channel measurements." IEEE Dataport (2018).

34.	Raed M. Shubair, and Hadeel Elayan. "Enhanced WSN localization of moving nodes using a robust hybrid TDOA-PF approach." In 2015 11th International Conference on Innovations in Information Technology (IIT), pp. 122-127. IEEE, 2015.

35.	M. I. AlHajri, N. T. Ali, and R. M. Shubair. "2.4 ghz indoor channel measurements data set." UCI Machine Learning Repository (2018).

36.	M. I. AlHajri, N. T. Ali, and R. M. Shubair. "2.4 ghz indoor channel measurements data set.” UCI Machine Learning Repository, 2018."

37.	WafaNjima, Marwa Chafii, and Raed M. Shubair. "Gan based data augmentation for indoor localization using labeled and unlabeled data." In 2021 International Balkan Conference on Communications and Networking (BalkanCom), pp. 36-39. IEEE, 2021.

38.	Mohamed I. AlHajri, Nazar T. Ali, and Raed M. Shubair. "A cascaded machine learning approach for indoor classification and localization using adaptive feature selection." AI for Emerging Verticals: Human-robot computing, sensing and networking (2020): 205.

39.	Mohamed I. AlHajri, Raed M. Shubair, and Marwa Chafii. "Indoor Localization Under Limited Measurements: A Cross-Environment Joint Semi-Supervised and Transfer Learning Approach." In 2021 IEEE 22nd International Workshop on Signal Processing Advances in Wireless Communications (SPAWC), pp. 266-270. IEEE, 2021.

40.	Raed M. Shubair and Hadeel Elayan. "In vivo wireless body communications: State-of-the-art and future directions." In 2015 Loughborough Antennas \& Propagation Conference (LAPC), pp. 1-5. IEEE, 2015.

41.	Hadeel Elayan, Raed M. Shubair, and Asimina Kiourti. "Wireless sensors for medical applications: Current status and future challenges." In 2017 11th European Conference on Antennas and Propagation (EUCAP), pp. 2478-2482. IEEE, 2017.

42.	Raed M. Shubair and Hadeel Elayan. "In vivo wireless body communications: State-of-the-art and future directions." In 2015 Loughborough Antennas \& Propagation Conference (LAPC), pp. 1-5. IEEE, 2015.

43.	Hadeel Elayan, Raed M. Shubair, and Asimina Kiourti. "Wireless sensors for medical applications: Current status and future challenges." In 2017 11th European Conference on Antennas and Propagation (EUCAP), pp. 2478-2482. IEEE, 2017.

44.	Hadeel Elayan, Raed M. Shubair, Josep Miquel Jornet, and Raj Mittra. "Multi-layer intrabody terahertz wave propagation model for nanobiosensing applications." Nano communication networks 14 (2017): 9-15.

45.	Hadeel Elayan, Pedram Johari, Raed M. Shubair, and Josep Miquel Jornet. "Photothermal modeling and analysis of intrabody terahertz nanoscale communication." IEEE transactions on nanobioscience 16, no. 8 (2017): 755-763.

46.	Rui Zhang, Ke Yang, Akram Alomainy, Qammer H. Abbasi, Khalid Qaraqe, and Raed M. Shubair. "Modelling of the terahertz communication channel for in-vivo nano-networks in the presence of noise." In 2016 16th Mediterranean Microwave Symposium (MMS), pp. 1-4. IEEE, 2016.

47.	Samar Elmeadawy and Raed M. Shubair. "6G wireless communications: Future technologies and research challenges." In 2019 international conference on electrical and computing technologies and applications (ICECTA), pp. 1-5. IEEE, 2019.

48.	Maryam AlNabooda, Raed M. Shubair, Nadeen R. Rishani, and GhadahAldabbagh. "Terahertz spectroscopy and imaging for the detection and identification of illicit drugs." 2017 Sensors networks smart and emerging technologies (SENSET) (2017): 1-4.

49.	Hadeel Elayan, Raed M. Shubair, and Asimina Kiourti. "On graphene-based THz plasmonic nano-antennas." In 2016 16th mediterranean microwave symposium (MMS), pp. 1-3. IEEE, 2016.

50.	Hadeel Elayan, Cesare Stefanini, Raed M. Shubair, and Josep Miquel Jornet. "End-to-end noise model for intra-body terahertz nanoscale communication." IEEE transactions on nanobioscience 17, no. 4 (2018): 464-473.

51.	Hadeel Elayan, Raed M. Shubair, Akram Alomainy, and Ke Yang. "In-vivo terahertz em channel characterization for nano-communications in wbans." In 2016 IEEE International Symposium on Antennas and Propagation (APSURSI), pp. 979-980. IEEE, 2016.

52.	Hadeel Elayan, Raed M. Shubair, and Josep M. Jornet. "Bio-electromagnetic thz propagation modeling for in-vivo wireless nanosensor networks." In 2017 11th European Conference on Antennas and Propagation (EuCAP), pp. 426-430. IEEE, 2017.

53.	Hadeel Elayan, and Raed M. Shubair. "On channel characterization in human body communication for medical monitoring systems." In 2016 17th International Symposium on Antenna Technology and Applied Electromagnetics (ANTEM), pp. 1-2. IEEE, 2016.

54.	Hadeel Elayan, Raed M. Shubair, and Nawaf Almoosa. "In vivo communication in wireless body area networks." In Information Innovation Technology in Smart Cities, pp. 273-287. Springer, Singapore, 2018.

55.	Hadeel Elayan, Raed M. Shubair, and Josep M. Jornet. "Characterising THz propagation and intrabody thermal absorption in iWNSNs." IET Microwaves, Antennas \& Propagation 12, no. 4 (2018): 525-532.

56.	Dana Bazazeh, Raed M. Shubair, and Wasim Q. Malik. "Biomarker discovery and validation for Parkinson's Disease: A machine learning approach." In 2016 International Conference on Bio-engineering for Smart Technologies (BioSMART), pp. 1-6. IEEE, 2016.

57.	Mayar Lotfy, Raed M. Shubair, Nassir Navab, and Shadi Albarqouni. "Investigation of focal loss in deep learning models for femur fractures classification." In 2019 International Conference on Electrical and Computing Technologies and Applications (ICECTA), pp. 1-4. IEEE, 2019.

58.	S. Elmeadawy, and R. M. Shubair. "Enabling technologies for 6G future wireless communications: Opportunities and challenges. arXiv 2020." arXiv preprint arXiv:2002.06068.

59.	Hadeel Elayan, Cesare Stefanini, Raed M. Shubair, and Josep M. Jornet. "Stochastic noise model for intra-body terahertz nanoscale communication." In Proceedings of the 5th ACM International Conference on Nanoscale Computing and Communication, pp. 1-6. 2018.

60.	Hadeel Elayan, Hadeel, and Raed M. Shubair. "Towards an Intelligent Deployment of Wireless Sensor Networks." In Information Innovation Technology in Smart Cities, pp. 235-250. Springer, Singapore, 2018.

61.	Abdul Karim Gizzini, Marwa Chafii, Ahmad Nimr, Raed M. Shubair, and Gerhard Fettweis. "Cnn aided weighted interpolation for channel estimation in vehicular communications." IEEE Transactions on Vehicular Technology 70, no. 12 (2021): 12796-12811.

62.	Nishtha Chopra, Mike Phipott, Akram Alomainy, Qammer H. Abbasi, Khalid Qaraqe, and Raed M. Shubair. "THz time domain characterization of human skin tissue for nano-electromagnetic communication." In 2016 16th Mediterranean Microwave Symposium (MMS), pp. 1-3. IEEE, 2016.

63.	Hadeel Elayan, Raed M. Shubair, Josep M. Jornet, Asimina Kiourti, and Raj Mittra. "Graphene-Based Spiral Nanoantenna for Intrabody Communication at Terahertz." In 2018 IEEE International Symposium on Antennas and Propagation \& USNC/URSI National Radio Science Meeting, pp. 799-800. IEEE, 2018.

64.	Taki Hasan Rafi, Raed M. Shubair, Faisal Farhan, Md Ziaul Hoque, and Farhan Mohd Quayyum. "Recent Advances in Computer-Aided Medical Diagnosis Using Machine Learning Algorithms with Optimization Techniques." IEEE Access (2021).

65.	Abdul Karim Gizzini, Marwa Chafii, Shahab Ehsanfar, and Raed M. Shubair. "Temporal Averaging LSTM-based Channel Estimation Scheme for IEEE 802.11 p Standard." arXiv preprint arXiv:2106.04829 (2021).

66.	Menna El Shorbagy, Raed M. Shubair, Mohamed I. AlHajri, and Nazih Khaddaj Mallat. "On the design of millimetre-wave antennas for 5G." In 2016 16th Mediterranean Microwave Symposium (MMS), pp. 1-4. IEEE, 2016.

67.	Ahmed A. Ibrahim,  JanMachac, and Raed M. Shubair. "Compact UWB MIMO antenna with pattern diversity and band rejection characteristics." Microwave and Optical Technology Letters 59, no. 6 (2017): 1460-1464.

68.	M. Saeed Khan, A-D. Capobianco, Sajid M. Asif, Adnan Iftikhar, Benjamin D. Braaten, and Raed M. Shubair. "A pattern reconfigurable printed patch antenna." In 2016 IEEE International Symposium on Antennas and Propagation (APSURSI), pp. 2149-2150. IEEE, 2016.

69.	M. Saeeed Khan, Adnan Iftikhar, Sajid M. Asif, Antonio‐Daniele Capobianco, and Benjamin D. Braaten. "A compact four elements UWB MIMO antenna with on‐demand WLAN rejection." Microwave and Optical Technology Letters 58, no. 2 (2016): 270-276.

70.	Muhammad Saeed Khan, Adnan Iftikhar, Antonio‐Daniele Capobianco, Raed M. Shubair, and Bilal Ijaz. "Pattern and frequency reconfiguration of patch antenna using PIN diodes." Microwave and Optical Technology Letters 59, no. 9 (2017): 2180-2185.

71.	Muhammad Saeed Khan, Adnan Iftikhar, Antonio‐Daniele Capobianco, Raed M. Shubair, and Bilal Ijaz. "Pattern and frequency reconfiguration of patch antenna using PIN diodes." Microwave and Optical Technology Letters 59, no. 9 (2017): 2180-2185.

72.	Muhammad Saeed Khan, Adnan Iftikhar, Raed M. Shubair, Antonio-D. Capobianco, Benjamin D. Braaten, and Dimitris E. Anagnostou. "Eight-element compact UWB-MIMO/diversity antenna with WLAN band rejection for 3G/4G/5G communications." IEEE Open Journal of Antennas and Propagation 1 (2020): 196-206.

73.	Amjad Omar, and Raed Shubair. "UWB coplanar waveguide-fed-coplanar strips spiral antenna." In 2016 10th European Conference on Antennas and Propagation (EuCAP), pp. 1-2. IEEE, 2016.

74.	Malak Y. ElSalamouny, and Raed M. Shubair. "Novel design of compact low-profile multi-band microstrip antennas for medical applications." In 2015 loughborough antennas \& propagation conference (LAPC), pp. 1-4. IEEE, 2015.

75.	Raed M. Shubair, Amna M. AlShamsi, Kinda Khalaf, and Asimina Kiourti. "Novel miniature wearable microstrip antennas for ISM-band biomedical telemetry." In 2015 Loughborough Antennas \& Propagation Conference (LAPC), pp. 1-4. IEEE, 2015.

76.	Ala Eldin Omer, George Shaker, Safieddin Safavi-Naeini, Georges Alquié, Frédérique Deshours, Hamid Kokabi, and Raed M. Shubair. "Non-invasive real-time monitoring of glucose level using novel microwave biosensor based on triple-pole CSRR." IEEE Transactions on Biomedical Circuits and Systems 14, no. 6 (2020): 1407-1420.

77.	Muhammad S. Khan, Syed A. Naqvi, Adnan Iftikhar, Sajid M. Asif, Adnan Fida, and Raed M. Shubair. "A WLAN band‐notched compact four element UWB MIMO antenna." International Journal of RF and Microwave Computer‐Aided Engineering 30, no. 9 (2020): e22282.

78.	Saad Alharbi, Raed M. Shubair, and Asimina Kiourti. "Flexible antennas for wearable applications: Recent advances and design challenges." (2018): 484-3.

79.	M. S. Khan, F. Rigobello, Bilal Ijaz, E. Autizi, A. D. Capobianco, R. Shubair, and S. A. Khan. "Compact 3‐D eight elements UWB‐MIMO array." Microwave and Optical Technology Letters 60, no. 8 (2018): 1967-1971.

80.	R. Karli, H. Ammor, R. M. Shubair, M. I. AlHajri, and A. Hakam. "Miniature Planar Ultra-Wide-Band Microstrip Patch Antenna for Breast Cancer Detection." Skin 1 (2016): 39.

81.	Mohammed S Al-Basheir, Raed M Shubai, and Sami M. Sharif. "Measurements and analysis for signal attenuation through date palm trees at 2.1 GHz frequency." (2006).

82.	Ala Eldin Omer, George Shaker, Safieddin Safavi-Naeini, Kieu Ngo, Raed M. Shubair, Georges Alquié, Frédérique Deshours, and Hamid Kokabi. "Multiple-cell microfluidic dielectric resonator for liquid sensing applications." IEEE Sensors Journal 21, no. 5 (2020): 6094-6104.

83.	Muhammad Saeed Khan, Adnan Iftikhar, Raed M. Shubair, Antonio-Daniele Capobianco, Sajid Mehmood Asif, Benjamin D. Braaten, and Dimitris E. Anagnostou. "Ultra-compact reconfigurable band reject UWB MIMO antenna with four radiators." Electronics 9, no. 4 (2020): 584.

84.	Amjad Omar, Maram Rashad, Maryam Al-Mulla, Hussain Attia, Shaimaa Naser, Nihad Dib, and Raed M. Shubair. "Compact design of UWB CPW-fed-patch antenna using the superformula." In 2016 5th International Conference on Electronic Devices, Systems and Applications (ICEDSA), pp. 1-4. IEEE, 2016.

85.	Muhammad S. Khan, Adnan Iftikhar, Raed M. Shubair, Antonio D. Capobianco, Benjamin D. Braaten, and Dimitris E. Anagnostou. "A four element, planar, compact UWB MIMO antenna with WLAN band rejection capabilities." Microwave and Optical Technology Letters 62, no. 10 (2020): 3124-3131.

86.	Ahmed A. Ibrahim, and Raed M. Shubair. "Reconfigurable band-notched UWB antenna for cognitive radio applications." In 2016 16th Mediterranean Microwave Symposium (MMS), pp. 1-4. IEEE, 2016.

87.	Hari Shankar Singh, SachinKalraiya, Manoj Kumar Meshram, and Raed M. Shubair. "Metamaterial inspired CPW‐fed compact antenna for ultrawide band applications." International Journal of RF and Microwave Computer‐Aided Engineering 29, no. 8 (2019): e21768.

88.	Omar Masood Khan, Qamar Ul Islam, Raed M. Shubair, and Asimina Kiourti. "Novel multiband Flamenco fractal antenna for wearable WBAN off-body communication applications." In 2018 International Applied Computational Electromagnetics Society Symposium (ACES), pp. 1-2. IEEE, 2018.

89.	Raed M. Shubair, Amer Salah, and Alaa K. Abbas. "Novel implantable miniaturized circular microstrip antenna for biomedical telemetry." In 2015 IEEE International Symposium on Antennas and Propagation \& USNC/URSI National Radio Science Meeting, pp. 947-948. IEEE, 2015.

90.	Sandip Ghosal, Arijit De, Ajay Chakrabarty, and Raed M. Shubair. "Characteristic mode analysis of slot loading in microstrip patch antenna." In 2018 IEEE International Symposium on Antennas and Propagation \& USNC/URSI National Radio Science Meeting, pp. 1523-1524. IEEE, 2018.

91.	Yazan Al-Alem, Ahmed A. Kishk, and Raed M. Shubair. "Enhanced wireless interchip communication performance using symmetrical layers and soft/hard surface concepts." IEEE Transactions on Microwave Theory and Techniques 68, no. 1 (2019): 39-50.

92.	Yazan Al-Alem, Ahmed A. Kishk, and Raed M. Shubair. "One-to-two wireless interchip communication link." IEEE Antennas and Wireless Propagation Letters 18, no. 11 (2019): 2375-2378.

93.	Yazan Al-Alem, Raed M. Shubair, and Ahmed Kishk. "Efficient on-chip antenna design based on symmetrical layers for multipath interference cancellation." In 2016 16th Mediterranean Microwave Symposium (MMS), pp. 1-3. IEEE, 2016.

94.	Nadeen R. Rishani, Raed M. Shubair, and GhadahAldabbagh. "On the design of wearable and epidermal antennas for emerging medical applications." In 2017 Sensors Networks Smart and Emerging Technologies (SENSET), pp. 1-4. IEEE, 2017.

95.	Asimina Kiourti, and Raed M. Shubair. "Implantable and ingestible sensors for wireless physiological monitoring: a review." In 2017 IEEE International Symposium on Antennas and Propagation \& USNC/URSI National Radio Science Meeting, pp. 1677-1678. IEEE, 2017.

96.	Yazan Al-Alem, Raed M. Shubair, and Ahmed Kishk. "Clock jitter correction circuit for high speed clock signals using delay units a nd time selection window." In 2016 16th Mediterranean Microwave Symposium (MMS), pp. 1-3. IEEE, 2016.

97.	Melissa Eugenia Diago-Mosquera, Alejandro Aragón-Zavala, Fidel Alejandro Rodríguez-Corbo, Mikel Celaya-Echarri, Raed M. Shubair, and Leyre Azpilicueta. "Tuning Selection Impact on Kriging-Aided In-Building Path Loss Modeling." IEEE Antennas and Wireless Propagation Letters 21, no. 1 (2021): 84-88.

98.	Mikel Celaya-Echarri, Leyre Azpilicueta, Fidel Alejandro Rodríguez-Corbo, Peio Lopez-Iturri, Victoria Ramos, Mohammad Alibakhshikenari, Raed M. Shubair, and Francisco Falcone. "Towards Environmental RF-EMF Assessment of mmWave High-Node Density Complex Heterogeneous Environments." Sensors 21, no. 24 (2021): 8419.

99.	Yazan Al-Alem, Ahmed A. Kishk, and Raed Shubair. "Wireless chip to chip communication link budget enhancement using hard/soft surfaces." In 2018 IEEE Global Conference on Signal and Information Processing (GlobalSIP), pp. 1013-1014. IEEE, 2018.

100.	Yazan Al-Alem, Yazan, Ahmed A. Kishk, and Raed M. Shubair. "Employing EBG in Wireless Inter-chip Communication Links: Design and Performance." In 2020 IEEE International Symposium on Antennas and Propagation and North American Radio Science Meeting, pp. 1303-1304. IEEE, 2020.

101. M. Alsabah et al., "6G Wireless Communications Networks: A Comprehensive Survey," in IEEE Access, vol. 9, pp. 148191-148243, 2021.

102. S. Elmeadawy and R. M. Shubair, "6G Wireless Communications: Future Technologies and Research Challenges," 2019 International Conference on Electrical and Computing Technologies and Applications (ICECTA), 2019, pp. 1-5.

103. Y. Xu, G. Gui, H. Gacanin and F. Adachi, "A Survey on Resource Allocation for 5G Heterogeneous Networks: Current Research, Future Trends, and Challenges," in IEEE Communications Surveys \& Tutorials, vol. 23, no. 2, pp. 668-695, Secondquarter 2021.

104. Ö. Özdogan, E. Björnson and E. G. Larsson, "Intelligent Reflecting Surfaces: Physics, Propagation, and Pathloss Modeling," in IEEE Wireless Communications Letters, vol. 9, no. 5, pp. 581-585, May 2020.

105. Y. Liu, X. Liu, X. Mu, T. Hou, J. Xu, M. Di Renzo, N. Al-Dhahir et al., "Reconfigurable Intelligent Surfaces: Principles and Opportunities," in IEEE Communications Surveys \& Tutorials, vol. 23, no. 3, pp. 1546-1577, thirdquarter 2021.

106. D. Zhao, H. Lu, Y. Wang and H. Sun, "Joint Passive Beamforming and User Association Optimization for IRS-assisted mmWave Systems," GLOBECOM 2020 - 2020 IEEE Global Communications Conference, 2020, pp. 1-6.

107. M. Mahbub and R. M. Shubair, "Intelligent Reflecting Surfaces in UAV-Assisted 6G Networks: An Approach for Enhanced Propagation and Spectral Characteristics," 2022 IEEE International IOT, Electronics and Mechatronics Conference, 2022, pp. 1-6.

108. S. Abeywickrama, C. You, R. Zhang and C. Yuen, "Channel Estimation for Intelligent Reflecting Surface Assisted Backscatter Communication," in IEEE Wireless Communications Letters, vol. 10, no. 11, pp. 2519-2523, Nov. 2021.

109. Y. Chen, Y. Wang, J. Zhang, P. Zhang and L. Hanzo, "Reconfigurable Intelligent Surface (RIS)-Aided Vehicular Networks: Their Protocols, Resource Allocation, and Performance," in IEEE Vehicular Technology Magazine, vol. 17, no. 2, pp. 26-36, June 2022.

110. M. Mahbub and R. M. Shubair, "Intelligent Reflecting Surfaces for Multi-Access Edge Computing in 6G Wireless Networks," 2022 IEEE International IOT, Electronics and Mechatronics Conference (IEMTRONICS), 2022, pp. 1-5.

111. Z. Sattar et al., "Full-Duplex Two-Tier Heterogeneous Network With Decoupled Access: Cell Association, Coverage, and Spectral Efficiency Analysis," in IEEE Access, vol. 8, pp. 172982-172995, 2020.

112. H. Hashida et al., "Mobility-Aware User Association Strategy for IRS-Aided mm-Wave Multibeam Transmission Towards 6G," in IEEE Journal on Selected Areas in Communications, vol. 40, no. 5, pp. 1667-1678, May 2022.

113. D. Zhao, H. Lu, Y. Wang, H. Sun and Y. Gui, "Joint Power Allocation and User Association Optimization for IRS-Assisted mmWave Systems," in IEEE Transactions on Wireless Communications, vol. 21, no. 1, pp. 577-590, Jan. 2022.

114. E. M. Taghavi, A. Alizadeh, N. Rajatheva, M. Vu and M. Latva-aho, "User Association in Millimeter Wave Cellular Networks with Intelligent Reflecting Surfaces," 2021 IEEE 93rd Vehicular Technology Conference (VTC2021-Spring), 2021, pp. 1-6.

115. W. Mei and R. Zhang, "Joint Base Station-IRS-User Association in Multi-IRS-Aided Wireless Network," GLOBECOM 2020 - 2020 IEEE Global Communications Conference, 2020, pp. 1-6.

116. W. Belaoura, K. Ghanem, M. Z. Shakir and M. O. Hasna, "Performance and User Association Optimization for UAV Relay-Assisted mm-Wave Massive MIMO Systems," in IEEE Access, vol. 10, pp. 49611-49624, 2022.

117. H. Munir, S. A. Hassan, H. Pervaiz, Q. Ni and L. Musavian, "User association in 5G heterogeneous networks exploiting multi-slope path loss model," 2017 2nd Workshop on Recent Trends in Telecommunications Research (RTTR), 2017, pp. 1-5.

118. T. Mir, L. Dai, Y. Yang, W. Shen and B. Wang, "Optimal FemtoCell Density for Maximizing Throughput in 5G Heterogeneous Networks under Outage Constraints," 2017 IEEE 86th Vehicular Technology Conference (VTC-Fall), Toronto, ON, Canada, 2017, pp. 1-5.

119. N. Hassan and X. Fernando, "Interference Mitigation and Dynamic User Association for Load Balancing in Heterogeneous Networks," in IEEE Transactions on Vehicular Technology, vol. 68, no. 8, pp. 7578-7592, Aug. 2019.

120. M. Mozaffari, W. Saad, M. Bennis and M. Debbah, "Optimal Transport Theory for Cell Association in UAV-Enabled Cellular Networks," in IEEE Communications Letters, vol. 21, no. 9, pp. 2053-2056, Sept. 2017.

121. W. Tang et al., "Wireless Communications With Reconfigurable Intelligent Surface: Path Loss Modeling and Experimental Measurement," in IEEE Transactions on Wireless Communications, vol. 20, no. 1, pp. 421-439.

122. M. O. Al-Kadri, Y. Deng, A. Aijaz and A. Nallanathan, "Full-Duplex Small Cells for Next Generation Heterogeneous Cellular Networks: A Case Study of Outage and Rate Coverage Analysis," in IEEE Access, vol. 5, pp. 8025-8038, 2017.

123. M. TEKBAŞ, A. TOKTAŞ and G. ÇAKIR, "Design of a Dual Polarized mmWave Horn Antenna Using Decoupled Microstrip Line Feeder," 2020 International Conference on Electrical Engineering (ICEE), 2020, pp. 1-4.

124. Ö. Özdogan, E. Björnson and E. G. Larsson, "Intelligent Reflecting Surfaces: Physics, Propagation, and Pathloss Modeling," in IEEE Wireless Communications Letters, vol. 9, no. 5, pp. 581-585, May 2020.

125. Y. Xie, X. Zhang, Q. Cui and Y. Lu, "User Association for Offloading in Heterogeneous Network Based on Matern Cluster Process," 2017 IEEE 85th Vehicular Technology Conference (VTC Spring), 2017, pp. 1-5.
}

\end{document}